\documentclass[12pt,preprint]{aastex}

\font\eightrm=cmr8

\def\be{\begin{equation}}
\def\ee{\end{equation}}
\def\beq{\begin{eqnarray}}
\def\eeq{\end{eqnarray}}

\def\ffrac#1#2{{\textstyle\frac{#1}{#2}}}
\def\C{{\cal C}}

\def\zb{{\bar z}}
\def\Khat{{\hat K}}
\def\Qb{{\bar Q}}
\def\zetab{{\bar\zeta}}

\def\det{{\rm det}}

\def\images{{\rm images}}

\def\ds{\displaystyle}
\def\Amin{A_{\rm min}}

\def\barz{\mathop{\bar{z}}\nolimits}
\def\barnu{\mathop{\bar{\nu}}\nolimits}
\def\barlam{\mathop{\bar{\lambda}}\nolimits}
\def\barzeta{\mathop{\bar{\zeta}}\nolimits}

\def\zs{z_{\rm s}}
\def\zl{z_{\rm l}}
\def\kmsMpc{{\rm km s}^{-1}{\rm Mpc}^{-1}}

\def\pd#1#2#3{{\ds \partial #1 \over \ds \partial #2}\Bigl|_#3}
\def\spose#1{\hbox to 0pt{#1\hss}}
\def\lta{\mathrel{\spose{\lower 3pt\hbox{$\sim$}} \raise
2.0pt\hbox{$<$}}}
\def\gta{\mathrel{\spose{\lower 3pt\hbox{$\sim$}} \raise
2.0pt\hbox{$>$}}}

\begin{document}
\title{Lensing Properties of Cored Galaxy Models}

\author{N. W. Evans}
\affil{Theoretical Physics, Department of Physics, 1 Keble Road,
Oxford, OX1 3NP, UK \\
w.evans1@physics.oxford.ac.uk}
\author{C. Hunter}
\affil{Department of Mathematics, Florida State University,
Tallahassee, Florida 32306-4510, USA \\
hunter@math.fsu.edu}

\begin{abstract}
A method is developed to evaluate the magnifications of the images of
galaxies with lensing potentials stratified on similar concentric
ellipses. In a quadruplet system, there are two even parity images,
two odd parity images, together with a de-magnified and usually
missing central image.  A simple contour integral is provided which
enables the sums of the magnifications of the even parity or the odd
parity images or the central image to be separately calculated without
explicit solution of the lens equation.  We find that the sums for
pairs of images generally vary considerably with the position of the
source, while the signed sums of the two pairs can be remarkably
uniform inside the tangential caustic in the absence of naked
cusps.

For a family of models in which the lensing potential is a power-law
of the elliptic radius, $\psi \propto (a^2 + x^2 + y^2
q^{-2})^{\beta/2}$, the number of visible images is found as a
function of flattening $q$, external shear $\gamma$ and core radius
$a$. The magnification of the central image depends on the size of the
core and the slope $\beta$ of the gravitational potential. It grows
strongly with the source offset if $\beta>1$, but weakly if $\beta<1$.
For typical source and lens redshifts, the missing central image leads
to strong constraints; the slope $\beta$ must be $\lta 1$ and the core
radius $a$ must be $\lta 300$ pc. The mass distribution in the lensing
galaxy population must be nearly cusped, and the cusp must be
isothermal or stronger. This is in good accord with the cuspy cores
seen in high resolution photometry of nearby, massive, early-type
galaxies, which typically have $\beta \approx 0.7$ (or surface density
falling like distance$^{-1.3}$) outside a break radius of a few
hundred parsecs. Cuspy cores by themselves can provide the explanation
of the missing central images. Dark matter at large radii may alter
the slope of the projected density; however, provided the slope
remains isothermal or steeper and the break radius remains small,
then the central image remains unobservable. The sensitivity of the
radio maps must be increased fifty-fold to find the central images in
abundance.
\end{abstract}

\keywords{gravitational lensing -- galaxies: structure -- galaxies:
kinematics and dynamics -- galaxies: halos}

\section{Introduction}

For smooth, non-singular lenses, it is well-known that the total
number of images is odd and that the number of even parity images
exceeds the number of odd parity images by one (e.g., Burke 1981;
Schneider, Ehlers \& Falco 1992, chap. 5). When the source lies within
both the tangential and radial caustics, it is lensed into 5 images. When
the source lies outside one caustic but inside the other, 
then it is lensed into 3 images. If the surface density at
the galaxy centre is cusped and that cusp is stronger than isothermal, 
then the situation changes.  There is no radial caustic and there
are either 4 or 2 images depending on whether the source lies inside or
outside the remaining tangential caustic (Evans \& Wilkinson 1998).

In fact, almost all the known sixty or so gravitational lens systems
are 2 or 4 image configurations (see Pospieszalka et al. 1999 for
details of the gravitational lensing database which maintains a list
of candidates). There are two systems in which the presence of a weak,
central image has been claimed.  APM08279+5255 is an ultraluminous
broad absorption line quasar. Originally, two images were detected
serendipitously by Irwin et al. (1998) in the optical. Subsequently,
Ibata et al. (1999) found convincing evidence for a central third
image using higher resolution infrared imaging. The source of
MG1131+0456 is a radio galaxy. One extended radio component is lensed
into a ring, another is lensed into two images. There seems to be a
weak central image at the center of the ring, and this may be the
third or fifth image of parts of the background radio source (Chen \&
Hewitt 1993); however, this could also be emission from the lensing
galaxy. The almost total absence of the central image from the known
lens systems sets strong constraints on the core radius and the
steepness of the lensing potential (e.g., Narayan, Blandford \&
Nityananda 1984, Wallington \& Narayan 1993, Rusin \& Ma 2000).

The aims of this paper are twofold.  On the theoretical side, our aim
is to find expressions for the sums of the magnifications of any
subset of the images produced by multiply lensed quasars or galaxies,
at least within the framework of a class of flexible and popular
models. We do this by extending the analysis of our earlier work
(Hunter \& Evans 2001, henceforth Paper I) to lensing potentials
stratified on similar concentric ellipse but this time with cores.  In
doublet (or quadruplet) systems, there are one (or two) even parity
images, one (or two) odd parity images, together with the missing
central image.  We provide contour integrals which find the sums of
the magnifications of the even parity images, the odd parity images
and the central image separately.  On the astrophysical side, our aim
is to set constraints on the cusp profile and core size of lensing
galaxies by requiring that the central image be too weak to be
detectable. For the 2 and 4 image systems, we investigate the
permissible core radius as a function of the steepness and the
flattening of the lensing potential and the external shear.

The paper is arranged as follows. Section 2 develops the contour
integral representation needed for lensing potentials that are
elliptically stratified. Section 3 specialises the analysis to models
in which the lensing potential is a power-law of the elliptic radius
combined with external shear of arbitrary orientation.  The conditions
for triple and quintuple imaging are found in terms of the depth and
shape of the potential, as well as the position of the source. The
contour integrals for the sums of the magnifications of the two even
parity, the two odd parity and the missing central image are
evaluated.  Section 4 uses Monte Carlo simulations to set limits on
the lensing potential by requiring the central image to be
unobservable.  Finally, our conclusions are given in Section 5.

\section{Contour Integrals for Observables}

This section develops contour integrals for computing the
magnifications of images assuming only that the lensing potential is
stratified on similar concentric ellipses.

\subsection{The Lens Equation}

In this paper, we always assume that the lensing potential $\psi$ is
stratified on similar concentric ellipses with constant axis ratio $q$
with $0<q\leq 1$ so that
\begin{equation}
\psi = \psi (\tau), \qquad\qquad \tau =x^2 + y^2q^{-2}.
\end{equation}
For a thin lens with potential $\psi$, the lens equation is (e.g.,
Schneider et al., section 5.1)
\begin{equation}
\label{eq:original}
\xi = x + \gamma_1 x + \gamma_2 y - 2x\psi^{\prime}(\tau),\qquad
\eta = y + \gamma_2 x - \gamma_1 y - {2y \over q^2}\psi^{\prime}(\tau),
\end{equation}
where ($\xi, \eta$) are Cartesian coordinates of the source. Here
$\gamma_1$ and $\gamma_2$ allow for a constant external shear in an
arbitrary direction. We always assume that the lens is not circularly
symmetric (that is, either $q\neq 1$ or $\gamma_1 \neq0$ or $\gamma_2
\neq0$).

Complex numbers often simplify calculations in lensing theory (e.g.,
Bourassa, Kantowski \& Norton 1973, Bourassa \& Kantowski 1975, Witt
1990, Rhie 1997). As in Paper I, we shall find it helpful to use
\begin{equation}
\label{eq:complexcoords}
\zeta = \xi + {\rm i} q \eta=|\zeta|e^{i\phi}, \qquad\qquad z = x + {\rm i}y/q,
\qquad\qquad \tau=z\zb.
\end{equation}
The lens equation (\ref{eq:original}) then becomes
\begin{equation}
\zeta = P_0 z +Q \zb -2z\psi^{\prime}(\tau),
\end{equation}
where
\begin{equation}
\label{eq:PQdefn}
P_0 =\ffrac{1}{2} [ 1+ q^2 + \gamma_1(1-q^2)],\qquad Q
=\ffrac{1}{2} [ 1- q^2 + \gamma_1(1+q^2) + 2{\rm i}q\gamma_2 ].
\end{equation}
It and its conjugate can be written in matrix form as
\begin{equation}
\label{eq:lenseq}
\left( \begin{array}{l}
\zeta \\
\zetab \\ \end{array} \right)
= \left( \begin{array}{lr}
P & Q \\
\Qb & P \\
\end{array} \right)
\left( \begin{array}{l}
z \\
\zb \\ \end{array} \right),
\end{equation}
where $P = P_0 -2\psi^{\prime}(\tau)$, from which it follows that
\begin{equation}
\label{eq:zzetarel}
\left( \begin{array}{l}
z \\
\zb \\ \end{array} \right)
= {1\over P^2 - |Q|^2}\left( \begin{array}{lr} P & -Q \\ -\Qb & P \\
\end{array} \right)
\left( \begin{array}{l}
\zeta \\
\zetab \\ \end{array} \right).
\end{equation}
When we form the real quantity $\tau = z\zb$ from eqn.~(\ref{eq:zzetarel}), 
we find that the solutions of the lens equation also satisfy the 
real equation
\begin{equation}
\label{eq:defnk}
K(\tau; \zeta, \zetab) \doteq {1 \over \tau} (P\zeta - Q \zetab )(P\zetab -\Qb
\zeta) - \left[ P^2 - |Q|^2\right]^2 =0.
\end{equation}
We designate $K(\tau; \zeta, \zetab) =0$ as {\it the imaging
equation}. Its real and positive solutions for $\tau$ provide the
image positions $z$ for a given source location $\zeta$.  Once a
solution for $\tau$ is found, the image positions are given by
equation (\ref{eq:zzetarel}).  Any solution of the original lens equation
(\ref{eq:original}) gives a solution of the imaging equation. However,
the imaging equation may have negative real or complex solutions
which do not correspond to true images.

\subsection{Sums of Magnifications}

The lens equation defines a map from the complex $(z,\zb)$--lens plane
to the $(\zeta, \zetab)$--source plane.  The Jacobian of this mapping
is
\begin{equation}
\label{eq:Jmatrix}
J = \left( \begin{array}{lr}
\pd{\zeta}{z}{\zb} &
\pd{\zeta}{\zb}{z} \\
\pd{\zetab}{z}{\zb} &
\pd{\zetab}{\zb}{z}
\end{array} \right)
= \left( \begin{array}{lr}
P + \tau P^{\prime}({\tau}) & Q + z^2P^{\prime}({\tau}) \\
\Qb + \zb^2P^{\prime}({\tau}) & P + \tau P^{\prime}({\tau}) \\
\end{array} \right),
\end{equation}
where the second form has been evaluated using eqn (\ref{eq:lenseq}).
The eigenvalues and eigenvectors of $J$ determine the distortion of
images. If the eigenvalues are both positive, the image is direct; if
the eigenvalues are both negative, the image is doubly inverted.
These are the even parity cases. If one eigenvalue is positive, the
other negative, then the image is inverted and the parity is odd.  The
reciprocal of the determinant of the Jacobian of the mapping from the
real $(x,y)$--lens plane to the $(\xi, \eta)$--source plane
corresponds physically to the signed magnification of an image
(Schneider et al. 1992, chapter 5).  With our choice
(\ref{eq:complexcoords}) of complex coordinates, we find
\begin{equation}
\label{eq:magperturb}
{q^2 \over \det J}\Biggl|_{x_i,y_i} = \mu_i p_i,
\end{equation}
where $\mu_i$ is the absolute value of the magnification and $p_i$ is
the parity of the image located at $(x_i, y_i)$. For certain
positions, $\det J$ vanishes and the magnification is infinite. These
are the critical points and lines. The caustics are the images of the
critical points and curves under the lens mapping (\ref{eq:original}).
Evaluating the determinant of (\ref{eq:Jmatrix}) gives
\begin{equation}
\label{eq:detJ}
\det J= P^2 - |Q|^2+(2\tau P -Q \zb^2 - \Qb z^2)P^{\prime}({\tau}).
\end{equation}
Differentiating the imaging equation (\ref{eq:defnk}) for $K$
partially with respect to $\tau$, holding $\zeta$ and $\zetab$ fixed,
and then using the lens equation (\ref{eq:lenseq}) to express $\zeta$
and $\zetab$ in terms of $z$ and $\zb$ gives
\begin{equation}
\label{eq:dKdtau}
\pd{K(\tau; \zeta, \zetab)}{\tau}{{\zeta,\zetab}}={-(P^2 - |Q|^2) \over \tau}
      \left[ P^2 - |Q|^2+(2\tau P -Q \zb^2 - \Qb z^2)P^{\prime}({\tau})\right].
\end{equation}
Consequently, we get the following compact expression for the signed
magnification of an image as
\begin{equation}
\label{eq:defmag}
\mu p={q^2 \over \det J} =  {-q^2 (P^2 - |Q|^2) \over
\tau \pd{K}{\tau}{{\zeta,\zetab}}}.
\end{equation}
Our contour integral representation relies on the special structure of
equation (\ref{eq:defmag}).  Images correspond to simple zeros of
$K(\tau;\zeta, \zetab)$ and $1/(\partial K/\partial \tau)$ is the
residue of $1/K$ at a simple zero.  We continue the right hand side of
equation (\ref{eq:defmag}) into the complex $\tau$-plane and write the
sum of the signed magnifications of the images as the contour integral
\begin{equation}
\label{eq:tautheorem}
\sum_{\images} \mu_i p_i = {-q^2 \over 2 \pi {\rm i}} \oint_\C
{d\tau \over \tau} { (P^2 - |Q|^2) \over K(\tau; \zeta, \zetab)}.
\end{equation}
Here, $\C$ is a contour in the complex $\tau$-plane which excludes
$\tau=0$ and encloses only the simple poles corresponding to whichever
visible images we wish to analyse.

Let us note that our analysis here shows how the methods first
developed in Paper I can do more than we achieved there. We here
exploit them in two new ways. We first relax the restriction that the
lensing potential be scale-free and allow it to have a core.  Cored
potentials produce either one, three, or five images depending on the
strength of the potential, the size of the core and the position of
the source. Second, we derive formulae for sums of signed
magnifications for separate image pairs, and for the central image.
Hence, the present results are more detailed and informative than
those of Paper I, most of which were for sums over the four bright
images, consisting of two pairs of opposite parity, and which
therefore partially cancel.

\section{Power-Law Galaxies with Cores}

We now specialize the analysis to the case of power-law galaxies with
a core radius $a$, which have the lensing potential
\begin{eqnarray}
\label{eq:lensing_pot}
\psi =\left\{ \begin{array}{ll} {\displaystyle A\over \displaystyle
\beta} (a^2+\tau)^{\beta/2} & \mbox{if $0 < \beta < 2$}, \\
\null&\null\\ {\displaystyle A\over \displaystyle 2} \ln (a^2+\tau)
& \mbox{ if $\beta =0$}. \end{array} \right.
\end{eqnarray}
These models were first studied by Blandford \& Kochanek (1989) in the
context of gravitational lensing (see also Kassiola \& Kovner 1993;
Witt 1996; Witt \& Mao 1997, 2000; Evans \& Wilkinson 1998 and Paper
I). They are the projections of three-dimensional power-law galaxies
familiar in galactic astronomy and dynamics (Evans 1993, 1994).  For
example, they have been used to model the nearby elliptical galaxy M32
(van der Marel et al. 1994), the inner parts of the Galactic bulge
(Evans \& de Zeeuw 1994), as well as the dark halo of our own Galaxy
in the interpretation of both the Sagittarius stream and the
microlensing results (Alcock et al. 1997; Ibata \& Lewis 1998). The
convergence (or surface density in units of the critical density) is
\begin{equation}
\label{eq:conv}
\kappa = {A\over 2q^2}\,{ a^2(1\!+\!q^2) + x^2 ( 1\!+\!q^2(\beta\!-\!1))
+ y^2(1\!+\!q^{-2}(\beta\!-\!1)) \over (a^2 + \tau)^{2 - \beta/2}}.
\end{equation}
It is easy to see that
\begin{equation}
\label{eq:lensing_pottwo}
2\psi^{\prime}(\tau)=A(a^2+\tau)^{-1/B},
\end{equation}
where the parameter $B=2 / (2-\beta)$ and has the range $1 \leq B <
\infty$. The positive parameter $A$, to which the magnitude of the
lensing potential is proportional, can be removed by a rescaling.
Specifically, we scale all lengths by $A^{B/2}$.  The net effect is to
set $A=1$ in eqs~(\ref{eq:lensing_pot})-(\ref{eq:lensing_pottwo}). The
dependence of our results on $A$, which is needed in applications, can
be recovered by multiplying all powers of $a$, $z$ and $\zeta$ and
their conjugates by the same powers of $A^{B/2}$.

\subsection{Images and Caustics}

\begin{figure}
\begin{center}
\plotone{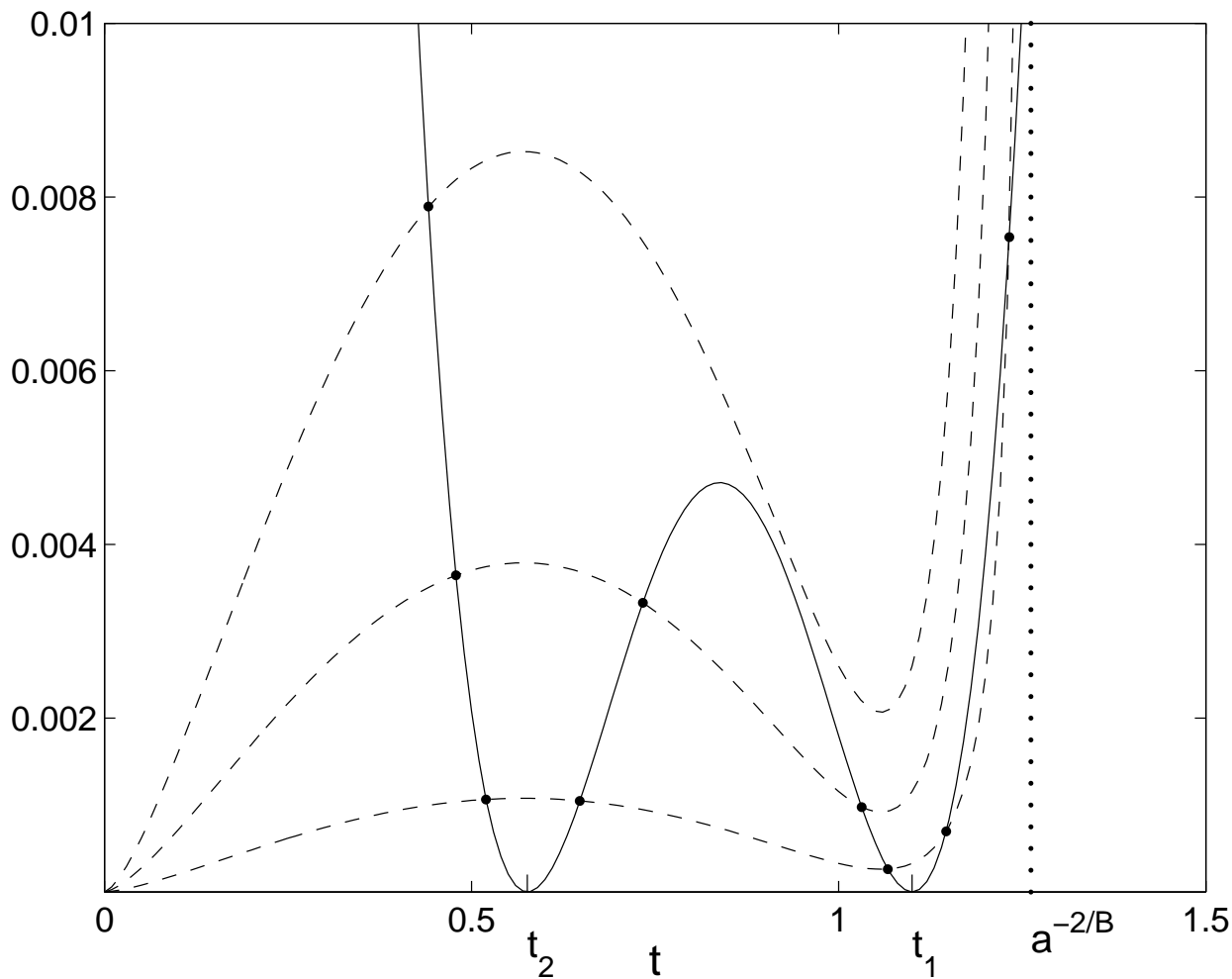}
\end{center}
\caption{Graphs of $K_2(t)/(1-a^2t^B)$ (dashed curves) and 
$-K_1(t)=(t-t_1)^2(t-t_2)^2$ (full curve) for the lens with $B=1.5$,
$q=0.8$, $a=0.84$, $\gamma_1=0.1$, and $\gamma_2=0$. The intersections of
the curves of different type correspond to images. The three dashed
curves, from the lowest up, are for the three source positions
$\zeta=|\zeta|\exp ({4\pi i/9})$ for $|\zeta| = .08$, $|\zeta| =
.15$, and $|\zeta| = .225$. These source positions give five, three,
and one images respectively.  Only for such a large core radius can a
figure be drawn in which all the intersections and the asymptote $t=a^{-2/B}$
are visible on the same scale.}
\label{fig:chrisone}
\end{figure}

\begin{figure}
\begin{center}
\plotone{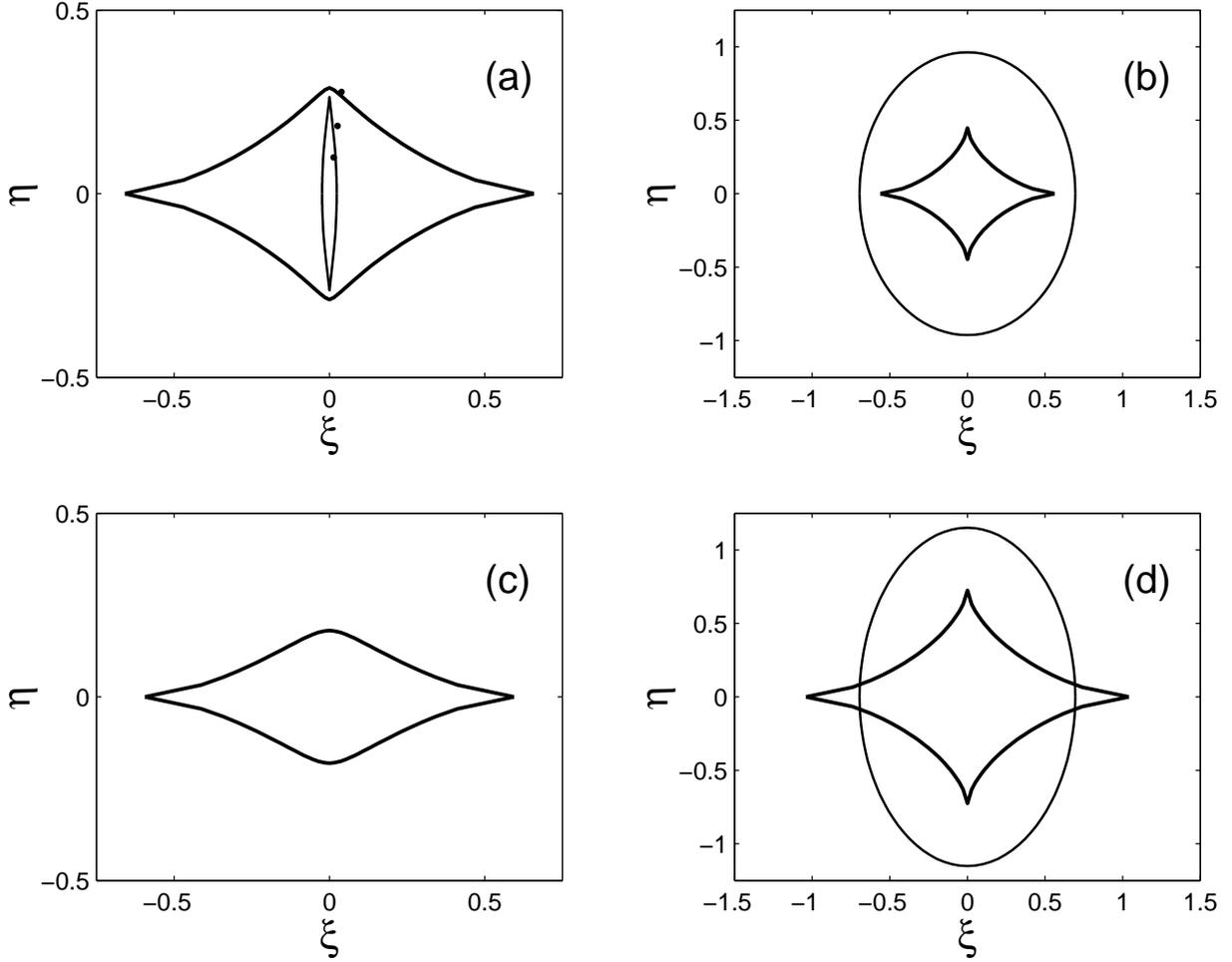}
\end{center}
\caption{The four different caustic configurations which can occur
with the cored elliptical potential (\ref{eq:lensing_pot}).  Case (a)
is for the double lips caustic for the lensing potential of
Fig.~\ref{fig:chrisone}. The dots show the three different source
positions, aligned and with increasing $|\zeta|$, for the dashed curves 
of Fig.~\ref{fig:chrisone}. Case (b) is for the
$B=2$, $q=0.8$, $a=0.1$, $\gamma_1=\gamma_2=0$ case for which
Fig.~\ref{fig:chrisfour} is plotted.  The single lips case (c) differs
from (a) only in having the larger core $a=1.0$ for which there is no
radial caustic.  Case (d) differs from (b) only in having the smaller
axis ratio $q=0.7$.  This model lies below the $B=2$ curve in 
Fig.~\ref{fig:christhree} and has naked cusps.}
\label{fig:christwo}
\end{figure}

In this section, we establish the conditions for the numbers and types
of images, and the forms of the caustics.
It is convenient to use the variable
$t=(a^2+\tau)^{-1/B}$. It reduces to the same variable $t$ used in
Paper I in the limit of no core ($a \to 0$). When there is a core, the
physically relevant range of $t$ is restricted to $0<t\leq
a^{-2/B}$. In terms of the variable $t$,
\begin{equation}
{1 \over \tau}={t^B \over 1-a^2t^B}, \qquad 2\psi^{\prime}(\tau)=t,
\qquad P=P_0-t.
\end{equation}
The imaging equation (\ref{eq:defnk}) requires the balance
\begin{equation}
\label{eq:tbalance}
{t^B (P\zeta - Q \zetab )(P\zetab -\Qb \zeta) \over 1-a^2t^B}
= (P^2-|Q|^2)^2 =(t-t_1)^2 (t-t_2)^2,
\end{equation}
for those real and positive values of $t$ at which the images occur.
The values of $t_1$ and $t_2$ depend only on the flattening
and the shear, and are defined by
\begin{equation}
\label{eq:tonetwodef}
t_1=P_0+|Q|, \qquad t_2=P_0-|Q|.
\end{equation}
It follows from the definitions (\ref{eq:PQdefn}) that $P_0>0$ and
$P^2_0-|Q|^2=q^2(1-\gamma^2_1-\gamma^2_2)>0$ provided that
$\gamma_1^2+\gamma_2^2<1$.  Hence, both $t_1$ and $t_2$ are positive
with $t_1>t_2$ and $t_2<1$.  Fig.~\ref{fig:chrisone} shows how image
positions can be found graphically by plotting separately the two
sides of equation (\ref{eq:tbalance}). The full curve represents the
right hand side of that equation. It depends only on the parameters of
the lens. The left hand side depends also on the position of the
source.  The three dashed curves display it for three different source
positions.  The lowest dashed curve intersects the full curve curve
five times and there are five images. When the source is sufficiently
close to the center of the galaxy, the left hand side remains small
until it rises to its asymptote at $a^{-2/B}$. A requirement for
quintuple imaging is that $t_1$ must lie to the left of this
asymptote, and hence that the core radius must satisfy $a<t_1^{-B/2}$.
Images disappear in pairs either when the two curves touch in the
$(t_2,t_1)$ interval, which corresponds to a crossing of the
tangential caustic, or when they touch in the $(t_1,a^{-2/B})$
interval, which corresponds to a crossing of the radial caustic.  If
the former happens first, then the three images that remain form a
``core triplet''; if the latter happens first, then the three images
form a ``naked cusp triplet'' (e.g., Kassiola \& Kovner 1993).  The
intermediate dashed curve in Fig.~\ref{fig:chrisone} represents a
naked cusp triplet case of a source which lies outside the radial
caustic but inside the tangential one.  The topmost dashed curve is
for a source which lies outside both caustics, and gives a single
image. There is now just a single image corresponding to a single
crossing on $(0,t_2)$.  The three different source positions are shown
in Fig.~\ref{fig:christwo}a.

The maximum number of images diminishes as the core radius increases.
If the core radius satisfies $a<t_1^{-B/2}$, then quintuple imaging is
possible, and there are two caustics.  For core radii in the range
$t^{-B/2}_1\leq a\leq t^{-B/2}_2$, the vertical asymptote of the
dashed curve in Fig.~\ref{fig:chrisone} lies between $t_1$ and $t_2$.
There are then three images for the source sufficiently close to the
center of the galaxy that it lies within the tangential caustic, and
there is no radial caustic.  For still larger core radii satisfying
$a\geq t^{-B/2}_2$, then only a single image can occur and there are
no caustics.  Appendix A justifies the statements concerning images
and caustics, and gives equations for determining the caustics, and an
approximate formula for the radial caustic for small $a$ and $B \leq 2$.  

\begin{figure}
\begin{center}
\plotone{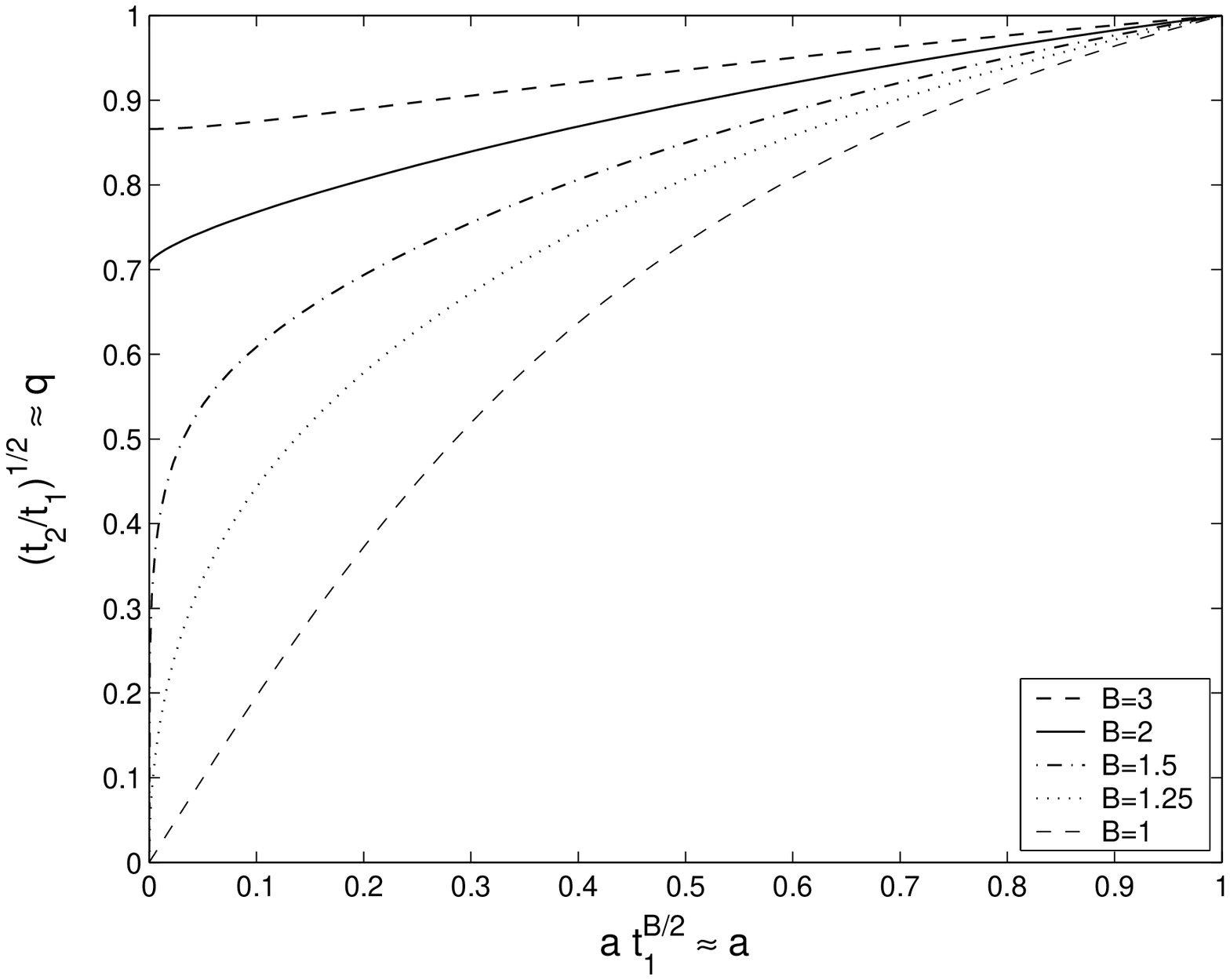}
\end{center}
\caption{The regions of parameter space in which the tangential caustic
has naked cusps are those which lie below the curves plotted. They grow
as $B$ increases. The values of $t_1$ and $t_2$ depend on the shear.
The approximations in the labels are those for negligible shear.
The radial caustic balloons in size as $a \to 0$ for $B<2$, as predicted
by equation (\ref{eq:radcausticsmalla}).}
\label{fig:christhree}
\end{figure}

Image positions and points on caustics must generally be determined
numerically. Cusps are an exception because they occur in pairs when
either $t=t_2$ or $t=t_1$ is a triple root of the imaging equation.
The sole caustic is tangential and of lips type for the range
$t^{-B/2}_1<a<t^{-B/2}_2$ as in Fig.~\ref{fig:christwo}c because it
has only the $t=t_2$ pair of cusps.  When $a<t^{-B/2}_1$ and quintuple
imaging is possible, there are two pairs of cusps and three possible
configurations of caustics.  Either both pairs of cusps lie on the
tangential caustic, which is of astroidal shape, or else both caustics
are lips-shaped with the radial caustic lying within the tangential
and oriented oppositely to it (Schneider et al. 1992, Section 8.6.1).
This double lips configuration occurs when
\begin{equation}
	\label{eq:twolips}
	1>a^2t^B_1>1-\frac{B}{2}
	\left(1-\frac{t_2}{t_1}\right),
\end{equation}
as in Fig.~\ref{fig:christwo}a. The $t=t_2$ pair of cusps are then
naked.  The stage at which those cusps become naked is described by
the coupled pair of equations (\ref{eq:nakedness}), which must be
solved numerically. The results are plotted in
Fig.~\ref{fig:christhree}. The tangential caustic has a pair of naked
cusps in the regions of parameter space below the curve for the
appropriate value of $B$. Above that curve, the tangential caustic has
four cusps and lies wholly within the radial caustic as in
Fig.~\ref{fig:christwo}b.  Below that curve, but where $t_2/t_1$ is
sufficiently large that the condition (\ref{eq:twolips}) is violated,
the two caustics intersect as in Fig.~\ref{fig:christwo}d, and two of
the four cusps of the tangential caustic are naked.

We note that our results extend previous calculations in the
literature. For example, Kassiola \& Kovner (1993) give the conditions
for multiple imaging in the isothermal case ($\beta =1$) in the
absence of shear ($\gamma_1 =0 = \gamma_2$), while Evans \& Wilkinson
(1998) give the results for the scale-free cases ($a=0$) with on-axis
shear only ($\gamma_2=0$).


\subsection{Magnifications of Pairs of Images}

Changing the variable in the contour integral (\ref{eq:tautheorem})
for the magnification to $t$ gives
\begin{equation}
\label{eq:ttheorem}
\sum_{\images} \mu_i p_i = {q^2 B \over 2 \pi {\rm i}} \oint_\C
{dt \over t} { P^2 - |Q|^2 \over (1-a^2t^B)K}.
\end{equation}
We write the denominator term as
\begin{equation}
\label{eq:newimaging}
(1-a^2t^B)K= \Khat_1 + K_2,
\end{equation}
where we have labelled its two parts as
\begin{eqnarray}
\label{eq:defnkone}
\Khat_1 &=&
-(1-a^2t^B)(P^2-|Q|^2)^2=-(1-a^2t^B)(t\!-\!t_1)^2(t\!-\!t_2)^2,
\nonumber \\
K_2 &=& t^B(P\zeta-Q\zetab)(P\zetab-\bar{Q}\zeta).
\end{eqnarray}
The component $K_2$ is the same as in Paper I, whereas $\Khat_1$
differs from the $K_1=-(t-t_1)^2(t-t_2)^2$ of Paper I only through 
its extra $(1-a^2t^B)$
factor, which is unity when there is no core.

In Paper I, we were interested primarily in sums over the four bright
images. However, the contour integral (\ref{eq:ttheorem}), like
the corresponding result in equation (23) of Paper I, applies
equally well to any subset of the images provided that the contour
$\C$ is chosen to enclose only the zeros of $(1-a^2t^B)K$ which
correspond to that subset.  If $a<t_2^{-B/2}$ for instance, we define
a contour $\C_2$ to be a simple closed contour in the complex
$t$--plane which encircles the two direct images of even parity. These
approach $t=t_2$ as $|\zeta| \to 0$, and so the contour $\C_2$ also
encloses $t_2$ which lies in between, but it excludes all the other
images, as well as $t=0$, $t=t_1$, $t=a^{-2/B}$ and any complex zeros
of $(1-a^2t^B)$.  Then, we obtain the formula
\begin{equation}
\label{eq:directmags}
	\sum\limits_{ {\hbox{\eightrm{2
direct}}}\atop{\hbox{\eightrm{images}}} } \mu_i p_i =
\sum\limits_{ {\hbox{\eightrm{2
direct}}}\atop{\hbox{\eightrm{images}}} } \mu_i =
{-q^2 B \over 2 \pi {\rm i}} \oint\nolimits_{\C_2}
{dt \over t(t-t_1)(t-t_2)(1-a^2t^B)\left(1+{K_2 \over \Khat_1}\right) }.
\end{equation}
We require that $\C_2$ keep a finite distance from $t_1$, $t_2$, and
every zero of $(1-a^2t^B)$. Then, with $|t|$ bounded above, and
$|t-t_1|$, $|t-t_2|$, and $|1-a^2t^B|$ bounded from below, it follows
that
\begin{eqnarray}
	\left|{K_2\over \Khat_1}\right|
	   & = & \left|{t^B\over 1-a^2t^B}\right|
\left|{|\lambda|^2\over(t-t_1)^2}+{|\nu|^2\over(t-t_2)^2}\right|\nonumber\\
		&\leq&\left|{t^B\over 1-a^2t^B}\right|
\left[{|\lambda|^2\over|t-t_1|^2}+{|\nu|^2\over|t-t_2|^2}\right],
\end{eqnarray}
and hence $|{K_2/\Khat_1}|<1$ on $\C_2$ for sufficiently small
$|\zeta|$.  We have here used the result
\begin{equation}
(P\zeta-Q\zetab)(P\zetab-\bar{Q}\zeta)=
[\lambda(t_2-t)+\nu(t_1-t)][\bar\lambda(t_2-t)+\bar\nu(t_1-t)]
=|\lambda|^2(t_2-t)^2+|\nu|^2(t_1-t)^2,
\end{equation}
where $\lambda$ and $\nu$ are the following two linear combinations
of the complex source coordinates
\begin{equation}
\label{eq:deflm}
           \lambda={1\over 2}\left(\zeta+{Q\zetab\over|Q|}\right),\qquad
\nu={1\over 2}\left(\zeta-{Q\zetab\over|Q|}\right),
\end{equation}
which have the property $|\lambda|^2+|\nu|^2=|\zeta|^2$.  With
$|{K_2/\Khat_1}|<1$, we can now expand the $(1+K_2/\Khat_1)^{-1}$ term
in the integral (\ref{eq:directmags}) as a geometric series to get
\begin{eqnarray}
	\sum\limits_{ {\hbox{\eightrm{2
direct}}}\atop{\hbox{\eightrm{images}}} }
	\mu_ip_i
&=&{-q^2B\over 2\pi {\rm i}}
\oint\nolimits_{\C_2}{dt\over t(t-t_1)(t-t_2)(1-a^2t^B)}\nonumber \\
& & - {q^2B\over 2\pi {\rm i}}\sum\limits^{\infty}_{j=1}\oint\nolimits_{\C_2}
	   {t^{Bj-1}[|\lambda|^2(t-t_2)^2+|\nu|^2(t-t_1)^2]^j \over
		 (t-t_1)^{2j+1}(t-t_2)^{2j+1}(1-a^2t^B)^{j+1}}dt.
\end{eqnarray}
The restrictions that we imposed earlier on $\C_2$ guarantee that this
series converges for sufficiently small $|\zeta|$. Every integral in
it can be evaluated from its residue at $t=t_2$ alone because that is
the only singularity within the contour $\C_2$. The leading order term
is the residue of the simple pole at $t=t_2$, which is positive
because both parities are even. The full expansion for the sum of the
two magnifications is
\begin{equation}
\label{eq:directseries}
	\sum\limits_{ {\hbox{\eightrm{2
direct}}}\atop{\hbox{\eightrm{images}}} }
	\mu_i
	={q^2B\over 2|Q|t_2(1-a^2t^B_2)}+ C_2(t_1,t_2,a^2;\lambda,\nu).
\end{equation}
Here, $C_2$ is a correction term which vanishes if the source is
exactly aligned with the center of the lensing galaxy. More generally,
it takes the form:
\begin{equation}
\label{eq:Ctwoseries}
C_2(t_1,t_2,a^2;\lambda,\nu) =
q^2B\sum\limits^{\infty}_{j=1}
	   \sum\limits^j_{m=0}{j\choose m}|\lambda|^{2m}|\nu|^{2j-2m}
	R_2(t_1,t_2,a^2;j,2m,2j-2m),\nonumber
\end{equation}
where we have defined
\begin{equation}
\label{eq:Rtwodefn}
	R_2(t_1,t_2,a^2;j,\ell,k)
	=-{1\over 2\pi {\rm i}}\oint\nolimits_{\C_2}
	 {t^{Bj-1}dt\over(t-t_1)^{\ell+1}(t-t_2)^{k+1}(1-a^2t^B)^{j+1}}.
\end{equation}
In fact, we can obtain closed form expressions for all of these
integrals without any further residue calculus by simple partial
differentiation because
\begin{eqnarray}
\label{eq:Rtwoderiv}
     R_2(t_1,t_2,a^2;j,\ell,k)
	&=&{1\over k!}{\partial^k\over\partial
t^k_2}R_2(t_1,t_2,a^2;j,\ell,0)\\
	&=&-{1\over k!}{\partial^k\over\partial t^k_2}   \left[
		{1\over(t_2-t_1)^{\ell+1}}
{t_2^{Bj-1}\over(1-a^2t^B_2)^j} \right].\nonumber
\end{eqnarray}

If the conditions for quintuple imaging are satisfied, then there is a
pair of inverted images which approach $t=t_1$ as $|\zeta|\to 0$. We
can perform a similar analysis for a simple closed contour $\C_1$ in
the complex $t$-plane which encloses the images and $t_1$, but not
$t=0$, $t=t_2$, $t=a^{-2/B}$ or any complex zero of $(1-a^2t^B)$.  The
result is that
\begin{equation}
\label{eq:invertedseries}
	\sum\limits_{ {\hbox{\eightrm{2
inverted}}}\atop{\hbox{\eightrm{images}}} }
	\mu_ip_i
	=-\sum\limits_{
{\hbox{\eightrm{2 inverted}}}\atop{\hbox{\eightrm{images}}} }
	    \mu_i
	={-q^2B\over 2|Q|t_1(1-a^2t^B_1)}
+C_1(t_1,t_2,a^2;\lambda,\nu),
\end{equation}
where the correction term $C_1$ is
\begin{equation}
C_1(t_1,t_2,a^2;\lambda,\nu) = q^2B\sum\limits^{\infty}_{j=1}
	\sum\limits^j_{m=0}{j\choose m}|\lambda|^{2m}|\nu|^{2j-2m}
	R_1(t_1,t_2,a^2;j,2m,2j-2m).
\end{equation}
Here, we have defined
\begin{eqnarray}
\label{eq:Ronederiv}
     R_1(t_1,t_2,a^2;j,\ell,k)
	&=&-{1\over 2\pi i}\oint\nolimits_{\C_1}
{t^{Bj-1}dt \over(t-t_1)^{\ell+1}(t-t_2)^{k+1}(1-a^2t^B)^{j+1}}\nonumber\\
	&=&{1\over\ell!}{\partial^{\ell}\over\partial t^{\ell}_1}
	   R_1(t_1,t_2,a^2;j,0,k)\\
	&=&-{1\over\ell!}{\partial^{\ell}\over\partial t^{\ell}_1}   \left[
	   {1\over(t_1-t_2)^{k+1}}
{t^{Bj-1}_1\over(1-a^2t^B_1)^j}\right]\nonumber\\
	&=&R_2(t_2,t_1,a^2;j,k,\ell).\nonumber
\end{eqnarray}
In fact, the coefficients $R_1(t_1,t_2,a^2;j,\ell,k)$ and
$R_2(t_1,t_2,a^2;j,\ell,k)$ can be expressed as finite sums, as
is demonstrated in Appendix B.

Neither series (\ref{eq:directseries}) nor (\ref{eq:invertedseries})
remains convergent at the tangential caustic where one of the direct
and one of the inverted images merge; the magnifications of the
merging images become infinite and one cannot then construct the
$\C_1$ and $\C_2$ contours.  However, the sum of series
(\ref{eq:directseries}) and (\ref{eq:invertedseries}) can remain
convergent at the tangential caustic because the infinities of the
signed magnifications of the two merging images cancel.  The series
(\ref{eq:invertedseries}) ceases to converge at the radial caustic
where an inverted and a doubly inverted image merge, but the series
(\ref{eq:directseries}) for the direct images converges there if, as
in Fig.~\ref{fig:christwo}a, it lies inside the tangential caustic.

\begin{figure}
\begin{center}
\plotone{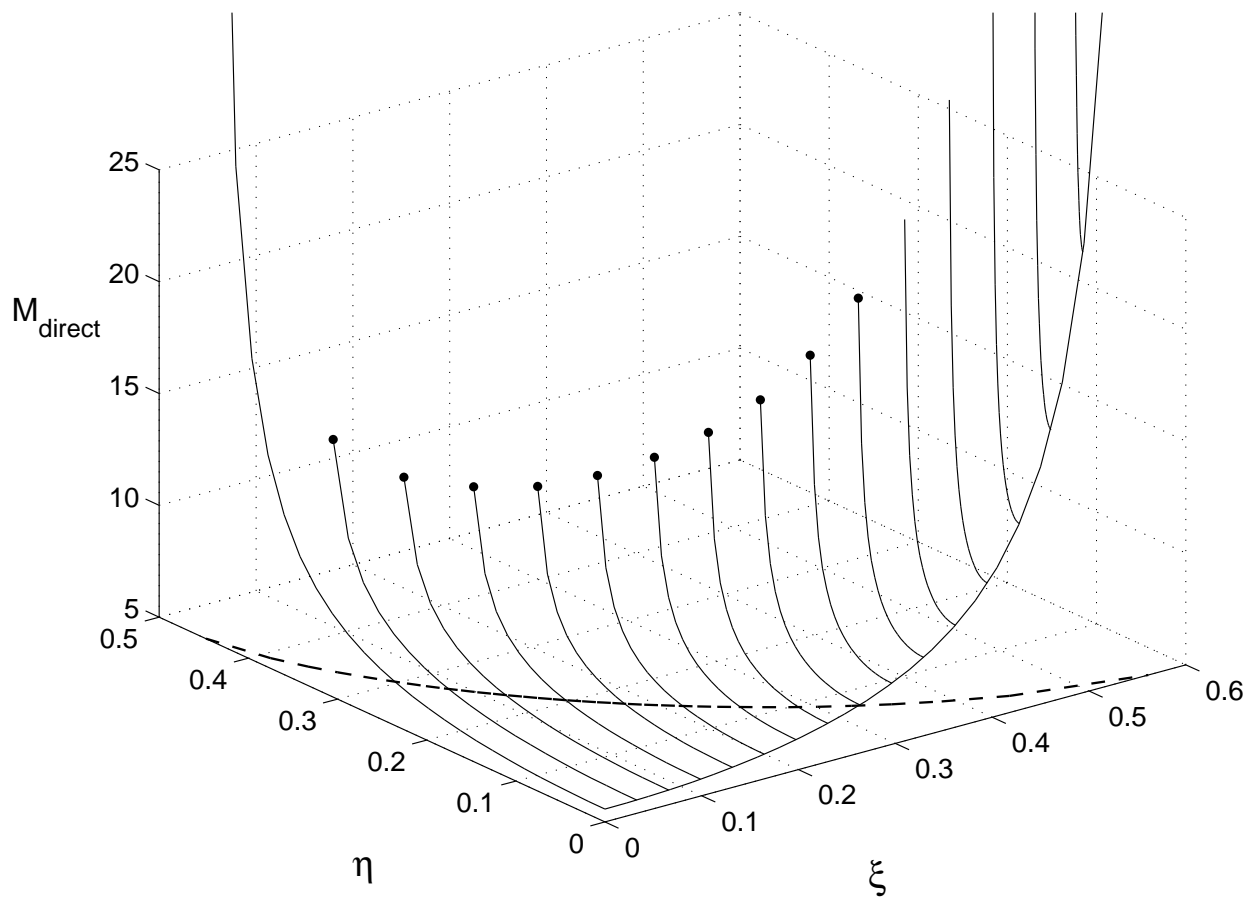}
\end{center}
\caption{The surface given by the sum of the magnifications of the two
direct images for $B=2$, $q=0.8$, $a=0.1$, $\gamma_1=\gamma_2=0$ over
the first quadrant of the $(\xi,\eta)$-plane inside the tangential
caustic.  That caustic is drawn as the dashed curve on the base of the
plot.  The full curves are the intersection of the surface with the
planes $\eta=0$, and evenly spaced planes $\xi={\rm const}$.  The
filled circles denote the points at which $\eta$ is $0.96$ of its
value at the caustic. Curves which do not end with filled circles
exceed 25 before getting that close to the caustic.}
\label{fig:chrisfour}
\end{figure}

Let us note some interesting special cases of the preceding formulae.
In the scale-free limit ($a=0$) when $B=1$ or $B=2$, then remarkably
the correction terms have the property that $C_1 - C_2 \equiv 0$. As
first discovered by Witt \& Mao (2000; see also Paper I), the sum of
the four signed magnifications is then an invariant completely
independent of the source position
\begin{equation}
	\sum\limits_{ {\hbox{\eightrm{4 images}}}} \mu_i p_i =
\frac{B}{ 1- \gamma_1^2 - \gamma_2^2}.
\label{eq:exact}
\end{equation}
Even though this result is not exact for other values of $B$, it is
often an extremely good approximation. If the core radius is
non-zero, then the sum of the signed magnifications becomes
\begin{equation}
	\sum\limits_{ {\hbox{\eightrm{4 images}}}} \mu_i p_i \approx
\frac{B}{ (1- \gamma_1^2 - \gamma_2^2)(1-a^2t_1^B)(1-a^2t_2^B)}
\left[ 1 - {a^2(t_1^{B+1} - t_2^{B+1}) \over t_1 - t_2} \right],
\label{eq:approx}
\end{equation}
and this approximation remains remarkably accurate inside the radial
caustic.  An example is the cored isothermal lens of
Fig.~\ref{fig:christwo}b. The sum of the four signed magnifications
is constant to within tenths of one percent over all of the region
inside the tangential caustic except very close to the cusps. However,
the smallness of the dependence of that sum on the source position is
due to a near-cancellation of the correction terms in the two image
sums.  Fig.~\ref{fig:chrisfour} shows how much the sum of the
magnifications of the two direct images varies over the same region.
Numerically, the leading coefficients in expansion
(\ref{eq:Ctwoseries}) for $C_2$ are $R_2(t_1,t_2,a^2;1,2,0)=13.77$ and
$R_2(t_1,t_2,a^2;1,0,2)=21.64$. The sums $R_1+R_2$ for the same sets
of indices are -0.01. Though numerical values vary, $R_1+R_2$ is
typically at least two orders of magnitude less than $R_2$ for lenses
with small cores and with an inner tangential caustic. In Section B.3,
we show how near-cancellation can occur more generally from the
contour integral formula (\ref{eq:tautheorem}), and is not a
peculiarity of power-law galaxies.  However, this behavior is not
universal. The sum of the four signed magnifications has large
variations over the region inside the inner radial caustic of
Fig.~\ref{fig:christwo}a, while the sum of the magnifications of the
two direct images changes little in the middle third of that region.
That middle third lies well inside the tangential caustic at which the
direct image sum becomes large.

The separate sums of the pairs contain more information than their
signed combination.  Physically speaking, it is the total
magnification (that is, the sum of the magnifications) which is more
interesting, but this varies considerably with source position. The
minimum magnification $\Amin$ in the scale-free limit $(a=0)$ is
\begin{equation}
\Amin = \sum\limits_{ {\hbox{\eightrm{4 images}}}} \mu_i = {B \over 1
- \gamma_1^2 - \gamma_2^2}{P_0 \over |Q|}
\end{equation}
This corresponds to the case when the source is aligned with the
center of the lensing galaxy. It is a good approximation only when
the source offset is small. For non-zero core radius, it becomes
\begin{equation}
\label{eq:Amincore}
\Amin = \sum\limits_{ {\hbox{\eightrm{4 images}}}} \mu_i = {B \over
1 - \gamma_1^2 - \gamma_2^2}{1\over 2|Q|}\left[ {t_2\over 1-a^2
t_1^B} + {t_1\over 1-a^2t_2^B} \right].
\end{equation}
In the next section, we discuss another nearly constant four-image sum
which reduces to the sum of the unsigned magnifications when the source
is aligned with the center of the lensing galaxy.

\subsection{Configuration Moments}
The contour integral method can also be used to calculate
configuration moments, that is, sums over the images of products of
the signed magnifications with position coordinates. Those sums are
obtained by adding the complex position coordinates from
(\ref{eq:zzetarel}) to the sum (\ref{eq:tautheorem}) to obtain
\begin{equation}
\label{eq:tautheorem_moments}
\sum_{\images} \mu_i p_i z_i^m \zb_i^n = {-q^2 \over 2 \pi {\rm i}} \oint_\C
{d\tau \over \tau}
{ (P\zeta - Q \zetab )^m(P\zetab -\Qb \zeta)^n
   \over K(\tau; \zeta, \zetab) (P^2 - |Q|^2)^{m+n-1}}.
\end{equation}
We expand for moments in the same manner as for the magnifications, with the
result that
\begin{equation}
\label{eq:dirconfigint}
         \sum\limits_{ {\hbox{\eightrm{2
direct}}}\atop{\hbox{\eightrm{images}}} }
         \mu_iz^m_i\barz^n_i = {-qB^2\over 2\pi i}\sum\limits^{\infty}_{j=0}
\oint\nolimits_{\C_2}{t^{Bj-1}[\lambda(t_2-t)+\nu(t_1-t)]^{m+j}
[\barlam(t_2-t)+\barnu(t_1-t)]^{n+j}dt
         \over[(t-t_1)(t-t_2)]^{2j+m+n+1}(1-a^2t^B)^{j+1}}.
\end{equation}
After the two numerator factors are expanded binomially, we see that
every integral is of the same form as (\ref{eq:Rtwodefn}),
and so, provided $m+j \geq 0$ and $n+j \geq 0$, we obtain
\begin{eqnarray}
\label{eq:dirconfigsum}
         \sum\limits_{ {\hbox{\eightrm{2
direct}}}\atop{\hbox{\eightrm{images}}} }\!\!\!\!\!\!
         &&\mu_iz^m_i\barz^n_i = (-1)^{m+n}q^2B\sum\limits^{\infty}_{j=0}
         \sum\limits^{2j+m+n}_{\ell=0}R_2(t_1,t_2,a^2;j,\ell,2j+m+n-\ell)
                                                         \nonumber\\
         &&\times\sum\limits^{\min(\ell,m+j)}_{k=\max(0,\ell-n-j)}{m+j\choose k}
         {n+j\choose\ell-k}\lambda^k\barlam^{\ell-k}\nu^{m+j-k}
         \barnu^{n+j+k-\ell}.
\end{eqnarray}
The corresponding sum for the indirect pair is
\begin{eqnarray}
\label{eq:invconfigsum}
         \sum\limits_{ {\hbox{\eightrm{2
inverted}}}\atop{\hbox{\eightrm{images}}} }\!\!\!\!\!\!
         &&\mu_iz^m_i\barz^n_i =
(-1)^{m+n+1}q^2B\sum\limits^{\infty}_{j=0}
         \sum\limits^{2j+m+n}_{\ell=0}R_1(t_1,t_2,a^2;j,\ell,2j+m+n-\ell)
                                                          \nonumber\\
         &&\times\sum\limits^{\min(\ell,m+j)}_{k=\max(0,\ell-n-j)}{m+j\choose k}
         {n+j\choose\ell-k}\lambda^k\barlam^{\ell-k}\nu^{m+j-k}
         \barnu^{n+j+k-\ell}.
\end{eqnarray}
In either case the same $R_1$ and $R_2$ functions as arise
with the magnifications, and which are evaluated in
Appendix B, are all that is needed. 

Whereas the present paper is mostly concerned with magnifications, our
modeling in Paper I made use of configuration moments too.  The
analysis of this paper, and especially that of Appendix Section B.3,
shows that four-image sums of moments may also be far more uniform
than the separate sums (\ref{eq:dirconfigsum}) and
(\ref{eq:invconfigsum}) because of cancellations.  However it is
possible to mitigate the cancellations which arise from signed
magnifications by using specially contrived configuration moments. As
an example, consider the zeroth order $m=1$, $n=-1$ moment [see eqn.
(48) of Paper I] in which the magnifications are weighted by the
complex exponential $e^{2i\Phi}$, where $z=|z|e^{i\Phi}$. The angle
$\Phi$ for each image is $\arctan(y_i/qx_i)$This factor exactly
compensates for the different signs of the magnifications when the
source offset is zero and the two kinds of images lie on perpendicular
lines through $O$. We obtain the $j=0$ terms directly from the
integral (\ref{eq:dirconfigint}) and its pair because the binomial
expansions used to obtain equations (\ref{eq:invconfigsum}) and
(\ref{eq:dirconfigsum}) are not valid when $n+j<0$. This gives us two
$R(t_1,t_2,a^2;0,0,0)$ integrals whose sum gives
\begin{equation}
\label{eq:newmagsum}
\sum\limits_{ {\hbox{\eightrm{4 images}}}} \mu_i p_ie^{2i\Phi_i}
\approx {-B \over (1 - \gamma_1^2 - \gamma_2^2)}
{1\over 2\Qb}\left[ {t_2\over 1-a^2
t_1^B} + {t_1\over 1-a^2t_2^B} \right],
\end{equation}
because $\lambda/\barlam=-\nu/\barnu=Q/|Q|$.  The sum
(\ref{eq:newmagsum}) is minus the sum in equation (\ref{eq:Amincore})
when there is no off-axis shear and $Q$ is real because $e^{2i\Phi}=1$
on the major $x$-axis where the inverted images initially lie and
$e^{2i\Phi}=-1$ on the minor $y$-axis.  Off-axis shear rotates the
configuration and $\bar{Q}/|Q|$ is then the complex factor which makes
$\sum\limits_{ {\hbox{\eightrm{4 images}}}}\mu_i p_i 
(\bar{Q}e^{2i\Phi}/|Q|)$ real. This four-image sum can remain nearly
constant as the source offset changes because the changing angular
positions of the images counteracts their changing magnitudes.  For
the cored isothermal lens for which Fig.~\ref{fig:christwo}b and
Fig.~\ref{fig:chrisfour} are plotted for example, the $m=1$, $n=-1$
moment of equation (\ref{eq:newmagsum}) displays the same near
constancy over the region inside the tangential caustic as does the
sum of the four signed magnifications. Estimates of $q$
and the orientation of the lens must be combined with observed
positions to evaluate the sum in (\ref{eq:newmagsum}), but the result 
should be real when multiplied by $\bar{Q}/|Q|$.

\subsection{The Magnification of the Central Image}

There is no fifth central image in the absence of a core for $B \leq 2$.
When there is a core, the central image is given by the root of the
imaging equation (\ref{eq:newimaging}) for which $t\to a^{-2/B}$ and
$\tau\to 0$ as $|\zeta|\to 0$. We can obtain formulae for its signed
magnification involving another set of $R$-functions, defined by
\begin{equation}
\label{eq:centralintegral}
	R_a(t_1,t_2,a^2;j,\ell,k)={-1\over 2\pi i}\oint\nolimits_{\C_a}
	{t^{Bj-1}dt\over(t-t_1)^{\ell+1}(t-t_2)^{k+1}(1-a^2t^B)^{j+1}}
\end{equation}
where $\C_a$ is a loop enclosing $t=a^{-2/B}$ but not $t_1$ or $t_2$
or $0$ or any other zeros of $(1-a^2t^B)$. These functions can be
calculated by evaluating residues at $t=a^{-2/B}$. This is a
simple pole when $j=0$, and we then get
\begin{equation}
	R_a(t_1,t_2,a^2;0,\ell,k)=
	{a^{2(\ell+k+2)/B}\over B[1-a^{2/B}t_1]^{\ell+1}[1-a^{2/B}t_2]^{k+1}}.
\end{equation}
The signed magnification of the central image is given by
\begin{equation}
\label{eq:central}
	\mu_a p={q^2a^{4/B}\over[1-a^{2/B}t_1][1-a^{2/B}t_2]} +
         C_a(t_1,t_2,a^2;\lambda,\nu).
\end{equation}
where the correction term $C_a(t_1,t_2,a^2;\lambda,\nu)$ has the form
\begin{equation}
\label{eq:centralC}
C_a(t_1,t_2,a^2;\lambda,\nu) = q^2B\sum\limits^{\infty}_{j=1}
	\sum\limits^j_{m=0}{j\choose m}|\lambda|^{2m}|\nu|^{2j-2m}
	R_a(t_1,t_2,a^2;j,2m,2j-2m).
\end{equation}
The lowest order term in~(\ref{eq:central}) is positive if
$a^{-2/B}>t_1$ when this image is doubly inverted, negative if
$t_1>a^{-2/B}>t_2$ and the image is simply inverted, and positive if
$t_2>a^{-2/B}$ and it is the only and direct image.

When $a$ is small, as in our applications, so that $a^{-2/B} \gg t_1$
and the central image occurs for a large value of $t$,
equations (\ref{eq:central}) and (\ref{eq:centralC}) can be approximated by
\begin{equation}
	\mu_a p=\mu_a= q^2a^{4/B}+q^2B\sum\limits^{\infty}_{j=1}
                       |\zeta|^{2j}R_a(0,0,a^2;j,2j,0).
\end{equation}
These special cases of integrals (\ref{eq:centralintegral}) 
can be evaluated by residues (change to $a^2t^B-1$ as integration variable) 
and give a series expansion for the magnification of the central image as
\begin{equation}
\label{eq:muasum}
        \mu_a= q^2a^{4/B}\left[1
                          +\sum\limits^{\infty}_{j=1}\alpha_j
                          (|\zeta|^2a^{2(2-B)/B})^j\right], \quad
        \alpha_j={\Gamma \left[{2(j+1) \over B}+1 \right]
                 \over j!\Gamma \left[{2(j+1) \over B}+1-j \right]}.
\end{equation}
The series is especially simple for $B=2$ when it is binomial and gives
\begin{equation}
\label{eq:muaBtwo}
        \mu_a= {q^2a^2 \over (1-|\zeta|^2)^2}.
\end{equation}
There are also two cases for which the series is hypergeometric and
can be summed explicitly. They are $B=1$, for which
\begin{equation}
\label{eq:muaBone}
        \mu_a= {4q^2a^4 \over \sqrt{1-4|\zeta|^2a^2}
                [1+\sqrt{1-4|\zeta|^2a^2}]^2},
\end{equation}
and $B=4$, for which
\begin{equation}
\label{eq:muaBfour}
        \mu_a= q^2a\left[{|\zeta|^2 \over a} + {1+|\zeta|^4/2a^2
                   \over \sqrt{1+|\zeta|^4/4a^2}} \right].
\end{equation}
The reason for the simplifications is that they are most easily obtained
directly from the basic magnification formula (\ref{eq:ttheorem}) 
after its integrand is approximated for large $t$ to give
\begin{equation}
	\mu_a={q^2B\over 2\pi i}\oint\nolimits_{\C_a}
	{dt\over t^3[t^{B-2}|\zeta|^2-1+a^2t^B]}.
\end{equation}
The root for the central image and its residue can be calculated 
directly without any need for series expansion for the three special
values of $B$. More generally, series expansion is needed. This analysis
makes it clear that the basic requirement for the approximation 
(\ref{eq:muasum}) to be valid is that $t$ be large at the central image.

\begin{figure}
\begin{center}
\plotone{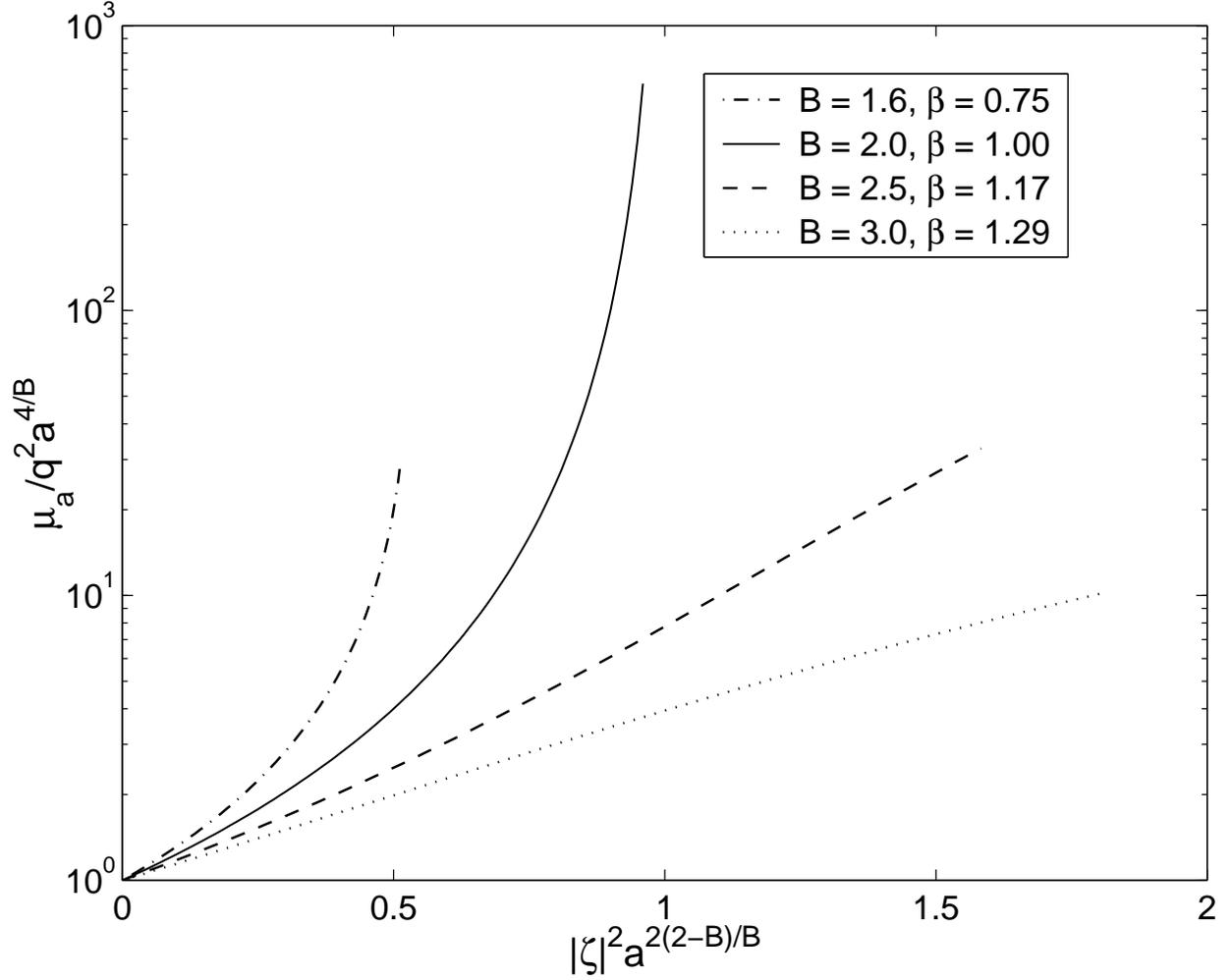}
\end{center}
\caption{Sums of the series (\ref{eq:muasum}) for the scaled magnification
of the central image as functions of its scaled argument out to $96\%$
of their radii of convergence. Those radii vary with $B$ as in
equation (\ref{eq:radconv}). The location of the radius of convergence
approximates the radial caustic for the top two curves with $B\leq 2$.
The near-linear growth with increasing source offset of the lower two 
curves for $B>2$ is a little less than, but growing towards,
the $B>2$ estimate $|\zeta|^{4/(B-2)}$ of equation (51) of Paper I.}
\label{fig:chrisfive}
\end{figure}

It is evident that the series (\ref{eq:muasum}) shows $\mu_a$ to be 
an increasing function of $|\zeta|^2a^{2(2-B)/B}$ for $B\leq 2$ because 
all its coefficients $\alpha_j$ are then positive. Numerical evaluation, 
as in Fig.~\ref{fig:chrisfive}, shows that $\mu_a$ remains an 
increasing function for larger $B$. Those plots extend to $96\%$ of the radius
of convergence. The series converge for
\begin{equation}
\label{eq:radconv}
      |\zeta|^2 < \left({B \over 2}\right)a^{2(B-2)/B}
                  \left|{B \over 2}-1\right|^{(2-B)/B},
\end{equation}
as follows from using Stirling's formula to approximate the gamma 
functions and standard convergence tests. Hence the range of usefulness 
of the series depends considerably on the value of $B$. 
For $B \leq 2$ and $a$ small, the radius of convergence is large, and 
the series converges out to the radial caustic as approximated by equation
(\ref{eq:radcausticsmalla}). The magnification of course becomes large as
the caustic, at $|\zeta|=1$ for $B=2$ and $|\zeta|=1/2a$ for $B=1$, 
is approached. For $B>2$ on the other hand, the radius of convergence 
is caused by the breakdown of the approximation that the central image 
occurs for large $t$ and, as with equation (\ref{eq:muaBfour}), 
$\mu_a$ remains finite as the condition (\ref{eq:radconv}) is violated. 
That radius of convergence in $|\zeta|$ is
proportional to the positive power $a^{1-2/B}$. Consequently, the
approximation (\ref{eq:muasum}) is useful for only a small part of
the region within the radial caustic when the core radius $a$ is small, 
and $\mu_a$, which is growing with source offset on a short length-scale, 
can continue to become considerably larger before the radial caustic is
approached.

Numerical tests shows that the approximation (\ref{eq:muasum}) works
well where the theory predicts that it will; over much of the region
within the radial caustic for $B \leq 2$, but only in limited central
regions for $B>2$.  It underestimates $\mu_a$ because of the neglect
of denominator terms such as those at the leading order in equation
(\ref{eq:central}). That underestimate can be by 10 \% or more for
$a=0.1$, but accuracy increases substantially for smaller $a$.

\begin{figure}
\begin{center}
\plotone{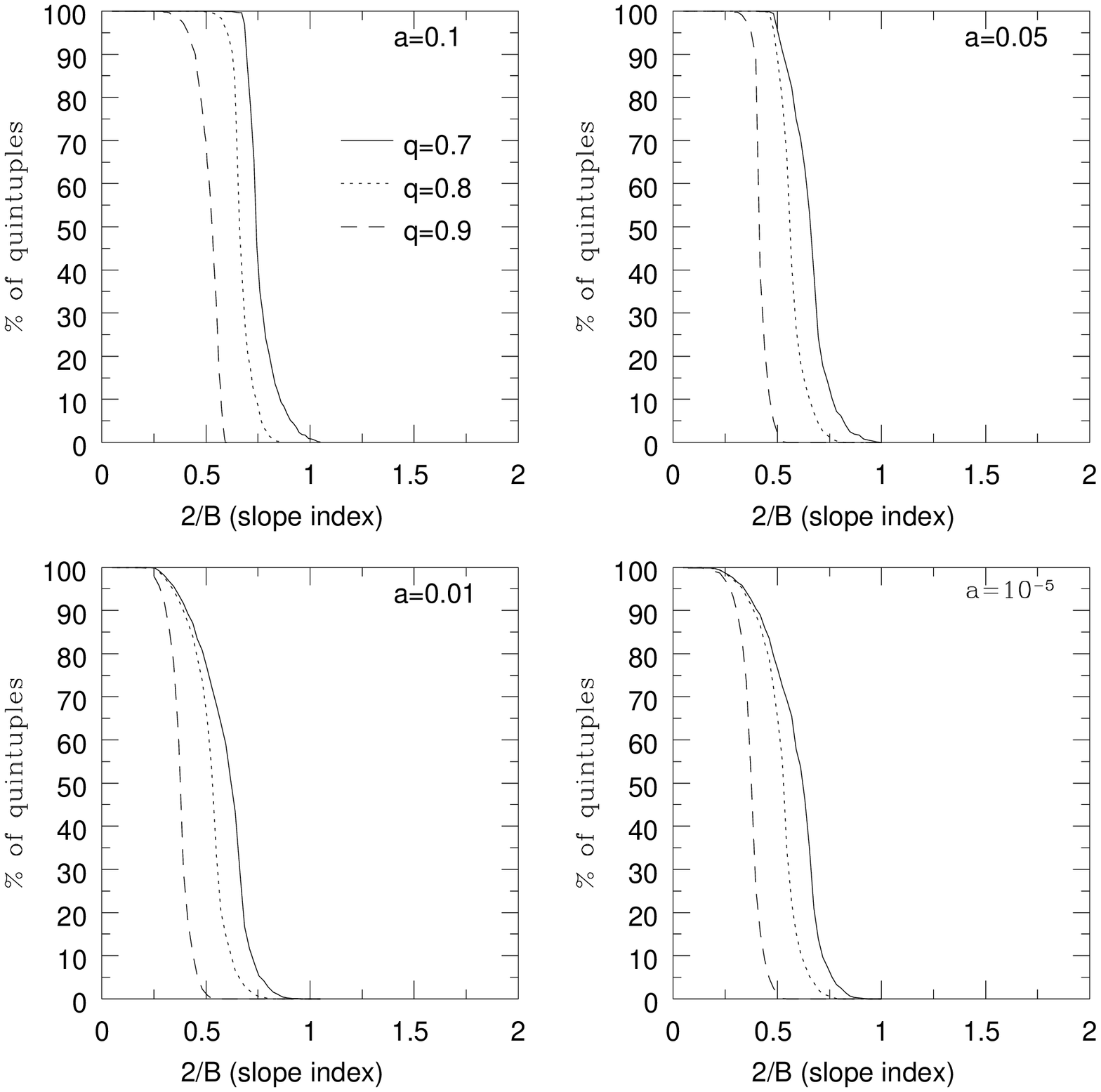}
\end{center}
\caption{The percentage of quintuplets with a visible central image is
plotted against the slope of the surface density.  If the model is
nearly singular, then $2/B >1$ for cusps steeper than isothermal,
while $2/B <1$ for cusps shallower than isothermal.  The four panels
show results for different values of the core radius $a$. There is no
fifth image when $a=0$ and $2/B \geq 1$. In each
panel, the full line denotes $q= 0.7$, the dotted line $q =0.8$ and
the dashed line $q=0.9$ models. The threshold is $1 \%$.}
\label{fig:wynzero}
\end{figure}
\begin{figure}
\begin{center}
\plotone{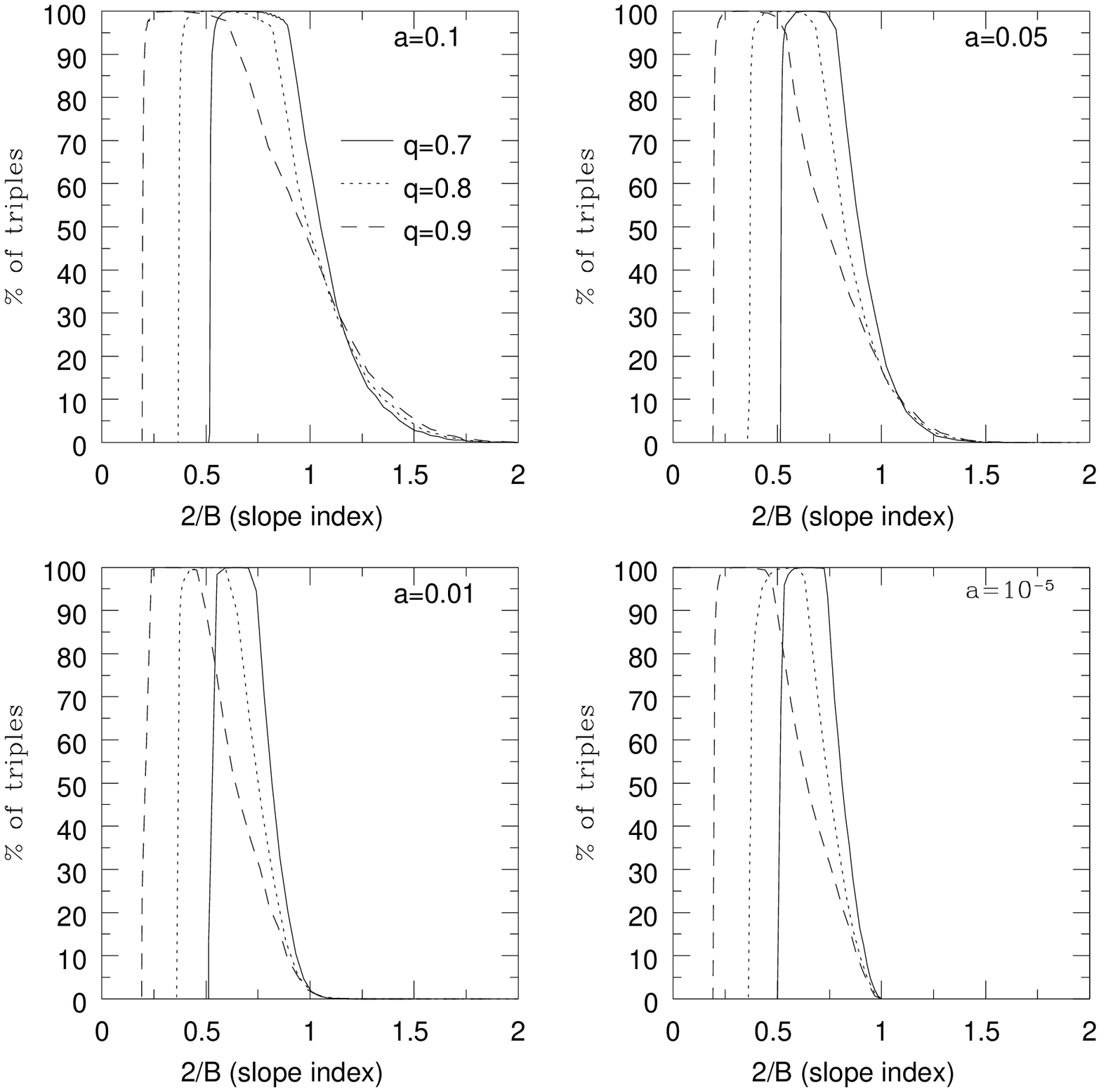}
\end{center}
\caption{The percentage of core triplets with a visible central image
is plotted against the slope of the surface density. The four panels
show results for different values of the core radius $a$. In each
panel, the full line denotes $q= 0.7$, the dotted line $q =0.8$ and
the dashed line $q=0.9$ models. The threshold is $1 \%$.}
\label{fig:wynone}
\end{figure}
\begin{figure}
\begin{center}
\plotone{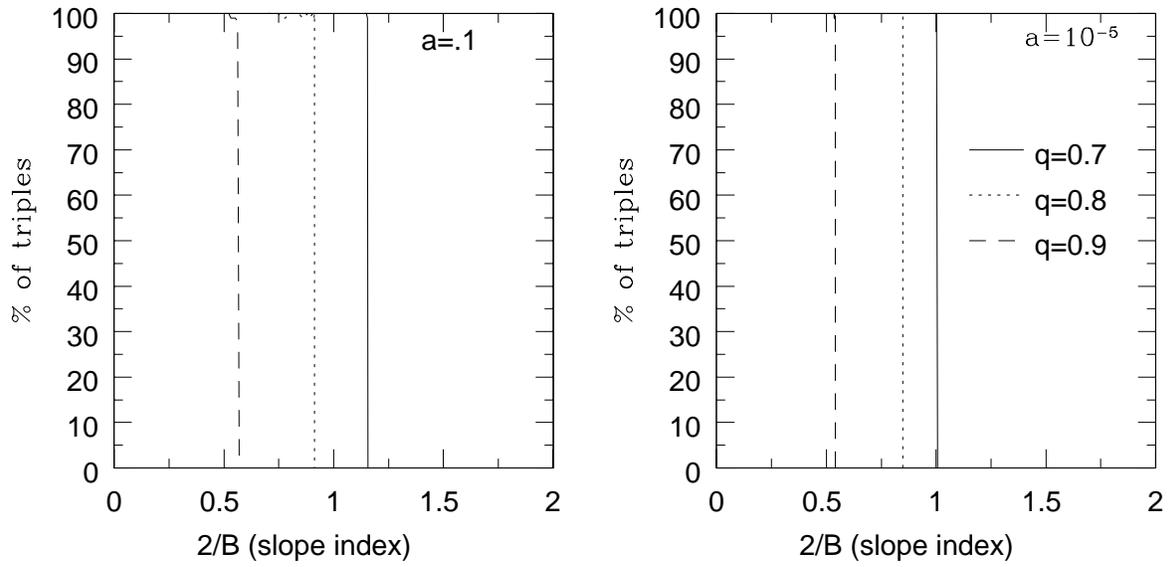}
\end{center}
\caption{The percentage of naked cusp triplets with a visible central
image is plotted against the slope of the surface density. The two
panels show results for different values of the core radius $a$. In
each panel, the full line denotes $q= 0.7$, the dotted line $q =0.8$
and the dashed line $q=0.9$ models. The threshold is $1 \%$.}
\label{fig:wyntwo}
\end{figure}

\section{Application: The Missing Central Images}

Here, we study the problem of missing central images from the
standpoint of the structure of the lensing galaxy. In optical lenses,
the experimental constraint is weak, as any central image can be
masked by emission from the lensing galaxy. Radio-loud lenses provide
much stronger constraints, as they have typically been probed with
high dynamic range radio maps. The {\it Cosmic Lens All-Sky Survey}
(CLASS) is the largest statistically homogeneous search for
gravitational lenses (e.g., Myers 1999).  The survey sample contains
$\sim 10^4$ flat-spectrum radio sources.  In the best cases, such as
B0218+357 (Biggs et al. 1999) and B1030+074 (Xanthopoulos et
al. 1998), the magnification ratio of the faintest to the brightest
image $r$ is constrained to $\lta 0.1 \%$ from the absence of a
detectable central image in the map.  As a typical detection limit in
the below calculations, we adopt $r \approx 1 \%$ for central
images. This is appropriate for radio lenses, but not for optical
lenses.

We perform Monte Carlo simulations using the power-law galaxies
(\ref{eq:lensing_pot}). Sources are placed randomly within the
outermost caustic for choices of $\beta$, $q$ and core radius $a$.
The imaging equation (\ref{eq:tbalance}) is solved numerically to find
the roots $t$ of the images. We then evaluate the magnifications using
(\ref{eq:defmag}), and work out the ratio of the brightest to the
faintest image.  Repeating this many times gives us the raw
probabilities of observing a central image.

Figs~\ref{fig:wynzero}-\ref{fig:wyntwo} show the probability of
observing a central image as a function of the slope of the projected
density ($2/B=2-\beta$) for the case of the quintuplets, the core
triplets and the naked cusp triplets respectively.  The panels show
how the raw probabilities depend on the core radius $a$, while the
different lines in each panel show different flattenings. The models
have a singular density profile if $a =0$, and are very nearly
singular when $a$ is small. We describe the parameter $2/B$ as the
slope index, as it controls the fall-off of the projected density. For
the isothermal cusp, $2/B =1$ and so the projected density $\kappa
\sim {\rm distance}^{-1}$. Models steeper than isothermal have $2/B
>1$, models shallower than isothermal have $2/B < 1$.

There are a number of things to notice in Fig.~\ref{fig:wynzero}.  As
the core radius $a$ is diminished, the r\'egime in which the central
image is visible shrinks. Visible central images occur only for $B>2$
when the central image is stronger as the simple estimate $\mu_a
\approx q^2a^{4/B}$ from \S 3.4 predicts.  The full curves for the
flattened ($q=.7$) models lie above the dashed curves for the rounder
($q=.9$) models.  This is due to a stronger dimming of the brightest
image with increasing flattening which outpaces the dimming of the
central image. All magnifications share a $q^2$-dependence which comes
from the basic magnification formula (\ref{eq:ttheorem}). The
brightest image is direct for vanishing shear and zero source offset,
and has magnification $\mu_1 \approx q^2B/(4|Q|t_2) =B/2(1-q^2)$.
Hence we estimate the ratio of the central image to the brightest
image at zero source offset to be $\mu_a/\mu_1 \approx
2q^2(1-q^2)Ba^{4/B}$. This ratio increases with increasing flattening.
For small $a$, the magnification of the central image does not vary
greatly with increasing source offset within the quintuplet region for
$B \leq 2$.  This region inside the tangential caustic is but a small
part of that within the radial caustic, and hence the abscissa of
Fig.~\ref{fig:chrisfive} varies little. Conversely, the brightest image
strengthens considerably with increasing source offset as it gets
closer to the tangential caustic.  The result is that the ratio
$\mu_a/\mu_1$ of central to brightest magnifications decreases with
increasing source offset for $B \leq 2$.  For $B>2$ on the other hand,
$\mu_a$ grows so strongly across the quintuplet region that it can
outpace the growth of $\mu_1$.  The net result is that the flatter the
potential, and the larger the core radius, then the greater is the
likelihood that the central image is bright enough to be visible.

Fig.~\ref{fig:wynone} shows the raw probability of observing a central
image for core triplets.  The behavior of the curves as a function of
slope index $2/B$ has the following explanation. If $2/B$ is too
large, then the central image is highly demagnified and so the raw
probability is vanishingly small. As the slope index $2/B$ diminishes,
the central image becomes brighter and the probability rises quickly
to $100 \%$. The brightest image is generally the remaining direct
one.  It dims with increasing source offset whereas the central image
brightens.  The tangential caustic also grows in size as $q$ decreases
for constant slope index.  That is the main reason why central images
are more visible with increased flattening for $B>2$; the average
source offset in the core triplet region between the tangential and
radial caustics is then larger, and hence $\mu_a$ is significantly
larger.  As $2/B$ decreases further, the astroidal tangential caustic
grows, and the area in the source plane generating core triplets
diminishes and eventually vanishes, as the tangential caustic becomes
larger than the radial caustic as in Fig.~\ref{fig:christwo}c.  The
smaller $q$ is, the sooner this happens and, as Fig.~\ref{fig:wynone}
shows, the size of the core has little effect on the stage at which
the core triplet region disappears. The second and third terms of the
inequality (\ref{eq:twolips}) become equal at that stage and, in the
absence of shear, $2/B$ is then $(1-q^2)/(1-a^2)$ (See Appendix A).
If the core radius $a \gta 0.01$, then even models steeper than
isothermal ($\beta<1$) can provide observable central images.  For
smaller core radii ($a < .01$), the visible central images are
confined to models less steep than isothermal. Notice that the
constraints on the maximum possible steepness of the lensing potential
provided by the missing image are stronger for doublets than
quadruplets once $a \lta .01$. For small core radii, the fifth image
in quintuplet systems is significantly more demagnified than the third
image in triplet systems.  However, for larger core radii, it is the
quadruplets that provide the stronger restriction.

Fig.~\ref{fig:wyntwo} shows the raw probability of observing a central
image for naked cusp triplets. Naked cusps are much less abundant for
elliptic potentials as opposed to elliptic densities (Kassiola \&
Kovner 1993). Almost as soon as naked cusps appear, all three images
are of roughly similar brightness and they are all detectable. Hence,
the raw probability of observing a central image shows a swift
transition from nearly $0\%$ to nearly $100 \%$ as soon as naked cups
become possible. The Monte Carlo results are consistent with the
transitions given in Fig.~\ref{fig:christhree}.  Equations
(\ref{eq:nakedness}) predict that the transitions occur at the values
$2/B = 0.565, \; 0.913, \; 1.157,$ for $a=0.1$ and $2/B = 0.542, \;
0.850, \; 1.004,$ for $a=10^{-5}$.  There are no strong observational
candidates for naked cusp triplets, and so we must conclude that
Fig.~\ref{fig:wyntwo} sets a firm lower limit on the slope index $2/B$.
This must be larger than the critical value which permits naked cusps
(given in Fig.~\ref{fig:christhree}), otherwise naked cusps would be
common.

\begin{figure}
\begin{center}
\plotone{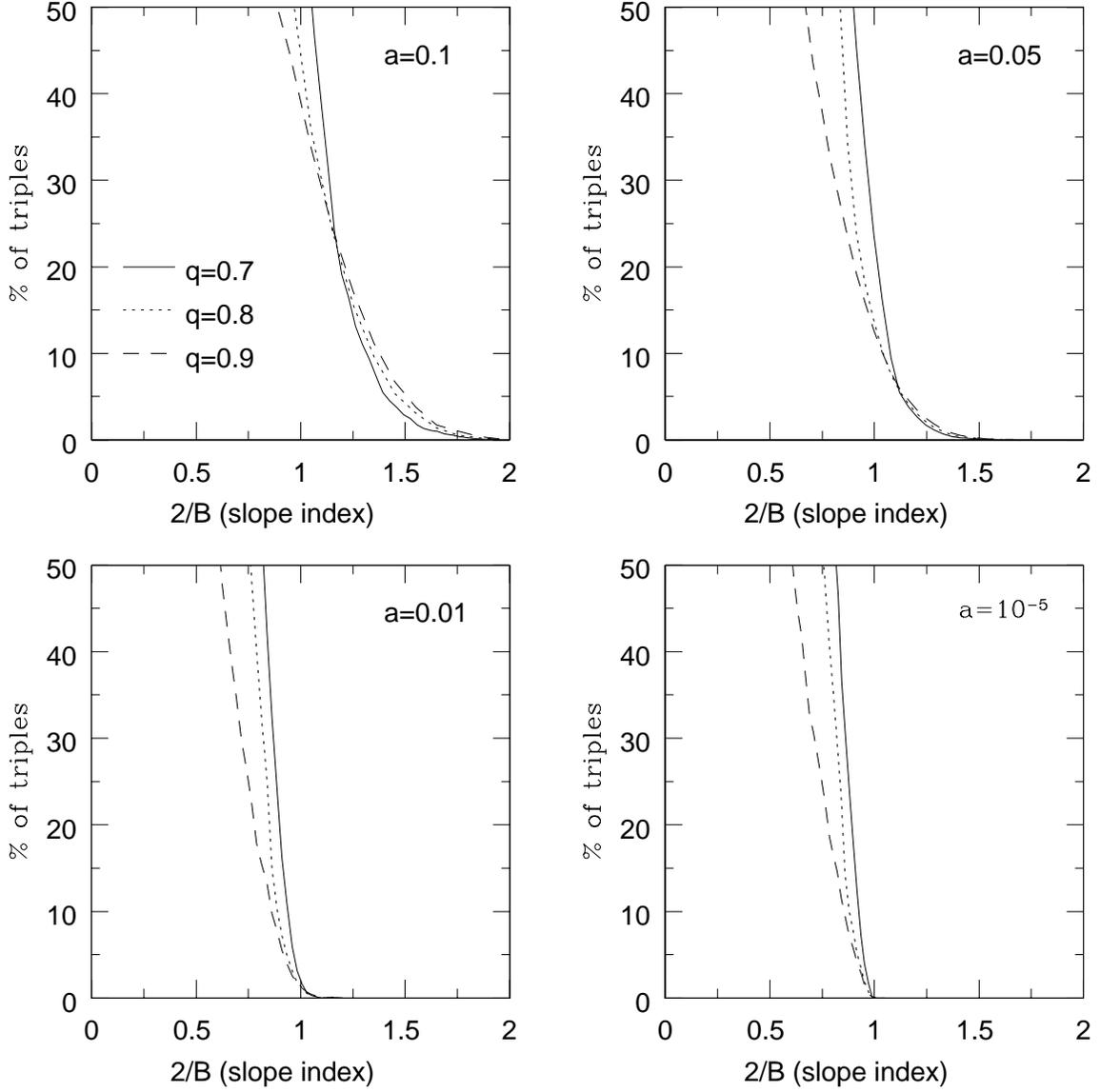}
\end{center}
\caption{The percentage of triplets with a visible central image is
plotted against the slope of the surface density. This figure
incorporates the amplification bias, i.e., the tendency of high
magnification configurations to be preferentially included in a
flux-limited sample. The threshold is $1 \%$. The increase of visibility
with increasing flattening is now due primarily to the increasing size of
the inner tangential caustic, which subtracts an area of weaker central
images from the triplet region.}
\label{fig:wynfour}
\end{figure}

\begin{figure}
\begin{center}
\plotone{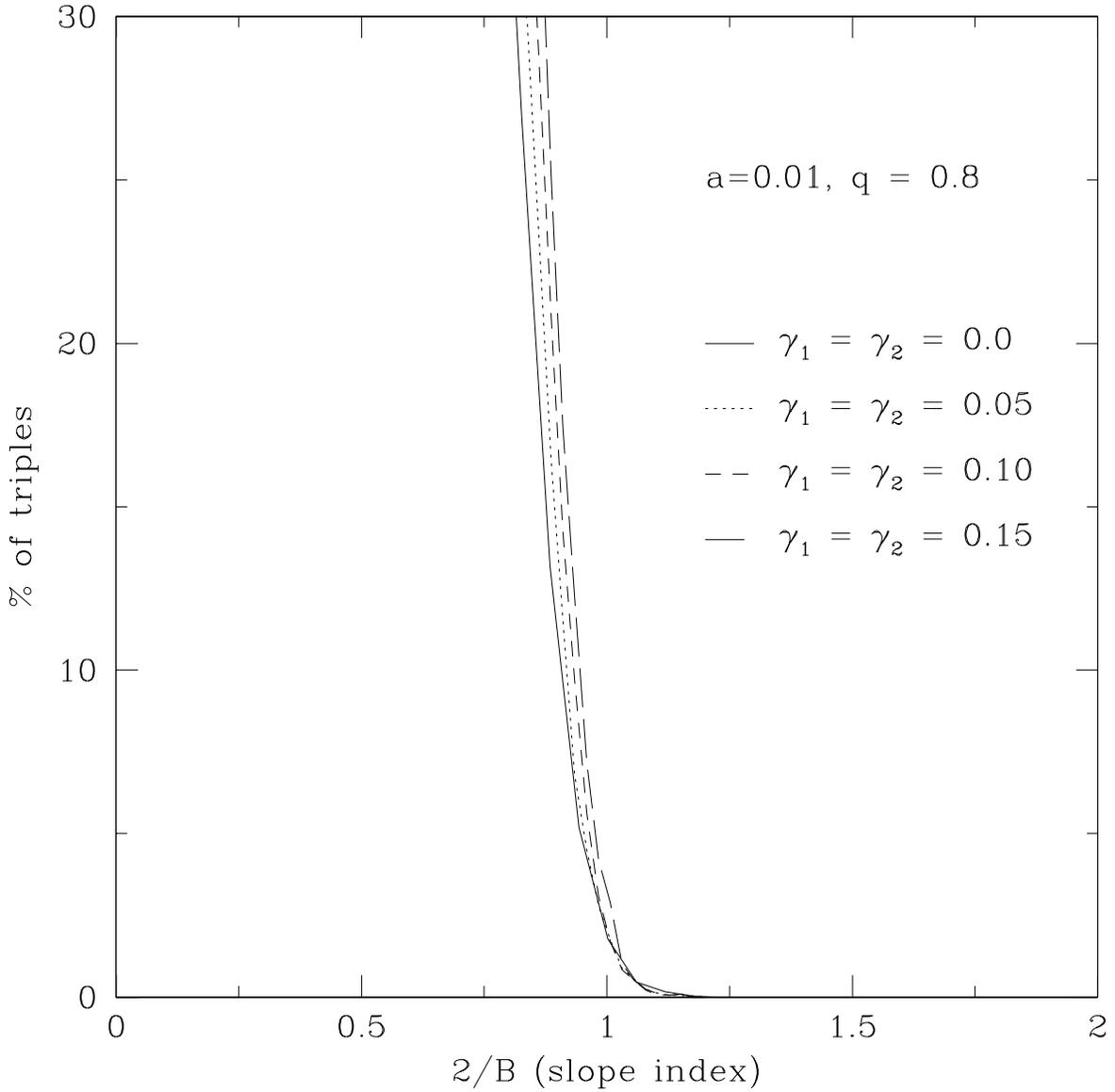}
\end{center}
\caption{This figure shows the effects of external shear. For models
with $a =0.01$, $q =0.8$, the percentage of triplets with a visible
central image is shown for a range of values of the shear components.
The threshold is $1 \%$.}
\label{fig:wynfive}
\end{figure}

To compare with data from surveys, we must allow for the amplification
bias. Lens systems with a high total magnification $\mu$ are
preferentially included in a flux-limited survey (e.g., Turner 1980,
Turner, Ostriker \& Gott 1984).  The flux distribution of the sources
in CLASS is well described by $dN/dS \propto S^{-2.1}$, where $N$ is
the number of sources with flux greater than $S$ (Rusin \& Tegmark
2001). For a flux limited sample, the probabilities that take into
account amplification bias are
\begin{equation}
P = \int_{\cal A} d\xi d \eta \left[ \mu(\xi, \eta) \right]^{1.1}
\end{equation}
where ${\cal A}$ denotes the area enclosed by the caustics in the
source plane for which the central image passes the threshold (e.g.,
Rusin \& Ma 2001). Fig.~\ref{fig:wynfour} is the analogue of
Fig.~\ref{fig:wynone} when amplification bias is taken into account.
Only the rightmost branch of the curve is plotted, as this is the most
relevant for constraining the core size and the slope index.  Notice
that the effects of the amplification bias cause only slight changes
in the shapes of the curves.  Once the core radius falls below $a
\approx 0.01$, then irrespective of the flattening, the lensing
potential is constrained to be at least as steep as isothermal to
ensure that the fraction of triplets with an observable central image
remains low.  As Fig.~\ref{fig:wynfive} shows, this conclusion remains
valid even in the presence of shear. Shear has little direct effect on the
magnitude of the central image as equation (\ref{eq:muasum}), which is
independent of shear, predicts. However, shears of the order of 0.1 cause the 
inner tangential caustic to grow significantly in size for the $2/B$
values at which the curves of Fig.~\ref{fig:wynfive} rise sharply, 
while the radial caustic merely tilts a little. The greater visibility
of the central image is again because the average source offset in the
diminished core triplet region has become larger, now as a result of shear. 
The total shear produced by internal misalignments, large-scale 
structure and neighboring galaxies is typically constrained by 
$|\gamma| \lta 0.3$.

\begin{figure}
\begin{center}
\plotone{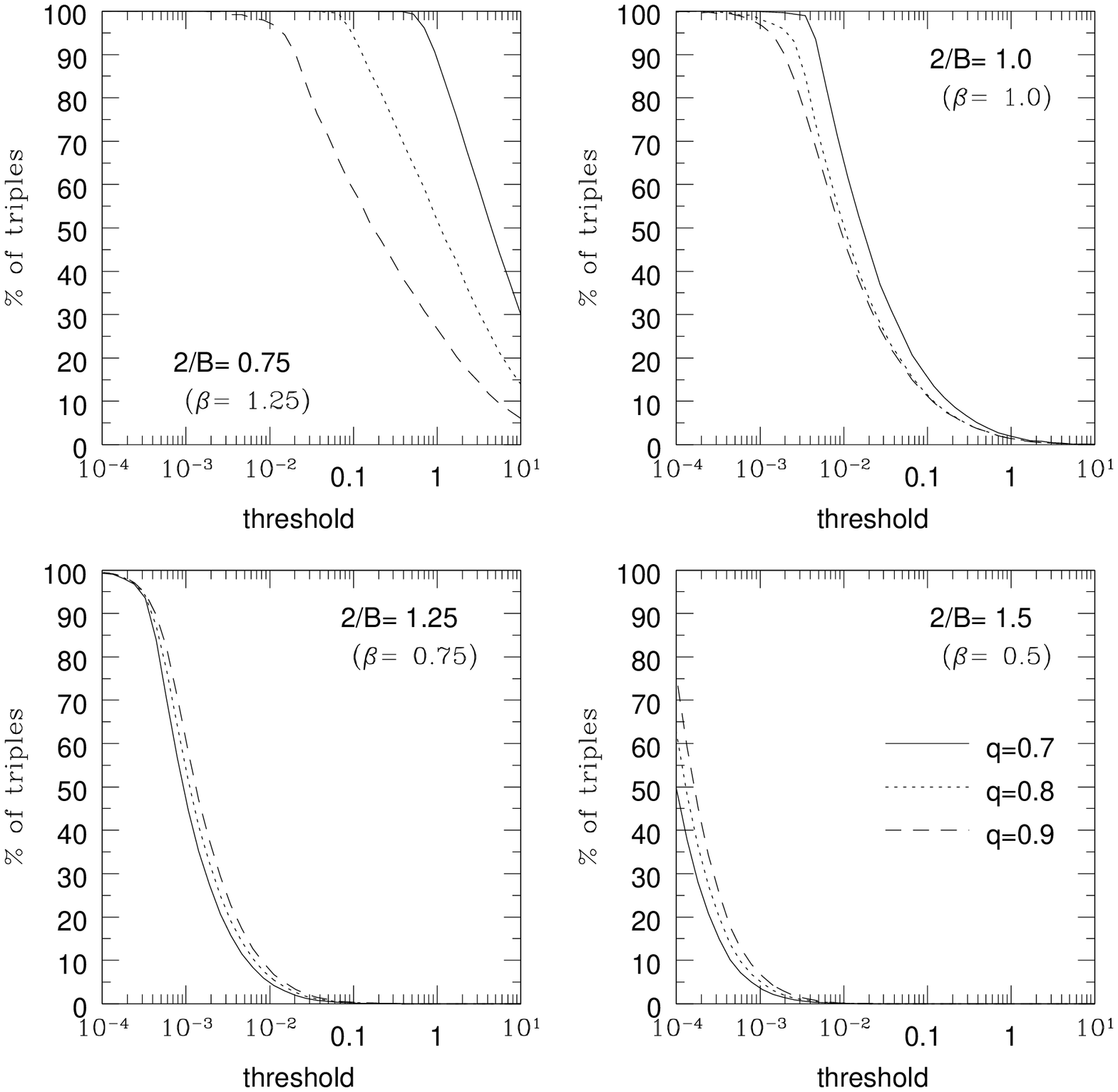}
\end{center}
\caption{The percentage of triplets with a visible central image is
plotted against the threshold (expressed as a percentage). The effects
of amplification bias are included.  The four panels show results for
different slope indices ($0.75, 1.0, 1.25$ and $1.5$).  In each panel,
the full line denotes $q= 0.7$, the dotted line $q =0.8$ and the
dashed line $q=0.9$ models. The core radius $a = 0.01$.}
\label{fig:wynsix}
\end{figure}

What restriction is implied on the core radius $a$ from the data on
radio lenses?  The CLASS survey found no triples, but 7
doublets. There is perhaps only 1 definite triple system
(APM08279+5255) out of a total of $\sim 50$ doublets and triplets on
Pospieszalka et al.'s gravitational lensing database (which contains
both radio and optical lenses). So, the probability of detecting a
triple with an observable central image is certainly very low. In this
paper, we take it to be $\lta 2 \%$ for radio lenses (i.e., at the $1
\%$ threshold). Using Fig.~\ref{fig:wynfour}, the top left panel
showing curves for $a = 0.1$ suggests that detectable central images
are common for isothermal (or nearly so) galaxies.  The smallest core
radius $a$ seemingly compatible with the missing central images is $a
\approx 0.01$ if the galaxy is isothermal. At a typical lens redshift
of $\zl \approx 0.5$ and a typical source redshift $\zs \approx 2.5$,
this corresponds to a physical size of $\approx 50$ pc (using formula
(11) of Young et al. (1980) and formula (2.4) of Wallington \& Narayan
(1993)).  The result holds good for a flat, matter-dominated
Friedman-Robertson-Walker universe with a Hubble constant $H_0 = 75
\,\kmsMpc$. Suppose instead we use the currently popular cosmological
model in which $\sim 70 \%$ of the mean energy density required to
make space flat is in the form of material for which gravity acts
repulsively and the remaining $\sim30\%$ is carried by collisionless
massive particles of some type. Then the physical size of the core
radii of isothermal galaxies is $\approx 70$ pc.  These results are
good for $2/B =1$.  From the top right panel of
Fig.~\ref{fig:wynfour}, the dimensionless core radius can be increased
to $a \approx 0.05$ if the galaxy has a larger slope index (say $2/B
\approx 1.3$, which we will argue shortly is appropriate for giant
ellipticals). In the same two cosmological models and with the same
assumptions as to typical source and lens redshifts, this gives
physical sizes of the core radius of $\approx 260$ pc and $\approx
350$ pc respectively, consistent with almost all the central images
being absent.

Most of the optical depth to strong lensing resides in giant
elliptical galaxies.  Faber et al. (1997) analyse {\it Hubble Space
Telescope} surface photometry of nearby ellipticals. They provide
convincing evidence that giant elliptical galaxies have cuspy cores
with steep outer power-law profiles and shallow inner profiles
separated by a break radius. This is in contrast to low luminosity
ellipticals, which have power-law surface brightness profiles.  There
are 26 giant ellipticals with cuspy cores presented in Table~2 of
Faber et al. The mean value of the asymptotic outer slope of the
surface brightness is $1.28$ with a standard deviation of $0.21$. This
corresponds to the slope index or $2/B$ in the notation of this
paper. The mean break radius is $330$ pc, which corresponds to only a
few tens of millarcseconds at a typical lens redshift of $\zl \approx
0.5$. The bright images of strong lenses are therefore probing the
steep outer part of the cuspy core profile.  The light profile is much
steeper than isothermal; however, the projected mass may be less
steep depending on the distribution of dark matter. Central images are
absent because break radii are small in cuspy core galaxies and so the
steep slope continues to small radii.  In fact, only 2 out of the 26
galaxies listed by Faber et al. (1997) have a sufficiently shallow
profile for the central image to stand any chance of being visible at
the $1 \%$ threshold.  This provides an explanation of why most
central images are unobservable at the current thresholds.
  
Other than a central black hole, there is little evidence for dark
matter inside the inner 10 kpc of early-type galaxies that cannot be
simply assigned to the stellar mass (e.g., Gerhard et al. 2001).
Beyond 10 kpc or so, there probably is dark matter, although there are
few hard facts on its distribution in ellipticals because of the
absence of tracers at large radii. The dynamical evidence refers to
the mass within spheres, whereas lensing is concerned with the mass
within cylinders. Dark matter at large radii may alter the slope of
the projected mass distribution in the inner few kpc. However,
provided the slope remains isothermal or steeper and the break radius
remains small, then the central image remains unobservable.
 
Fig.~\ref{fig:wynsix} shows the percentage of triplets with a visible
central image as a function of the threshold for four different slope
indices. Notice that for models with slopes steeper than isothermal,
flattening typically makes the central image less visible, the reverse
of the case when the slope is weak. The latter case shows the same
trends as in the $a=0.01$ panel of Fig.~\ref{fig:wynone}. The radial
caustic is large when $B<2$ and the tangential caustic occupies a much
smaller part of the space within it. Both $\mu_a$ and $\mu_1$ now
decrease as $q$ decreases, but $\mu_a$ decreases faster to make the
central image less visible.  Fig.~\ref{fig:wynsix} shows us that the
fraction of detectable images is a strong function of the threshold.
It enables us to predict the threshold required to find the missing
central images.  Using the third panel of Fig.~\ref{fig:wynsix} as
typical of the outer parts of cuspy core profiles, we see that the
threshold has to be $\sim 0.001 \%$ for the central image to be
detectable in half the triplet systems. Even the most sensitive CLASS
radio map (for B0218+357) probes only to a threshold of $0.05 \%$, so
this provides an explanation of why the central images have so far
remained missing despite deeper searches.

\section{Conclusions}

We have shown how contour integration can ease the evaluation of the
magnifications of images. This work extends the ideas presented in
Hunter \& Evans (2001) in two significant ways. First, our earlier
analysis was restricted to scale-free power-law potentials.  We have
now generalized it to cover all elliptically stratified potentials.
Second, we have obtained separate formulas for sums of the two direct
images, for sums of the two inverted images, and for the magnification
of the central image. Previously, we found only formulas for sums over
all four images weighted with the signed magnifications.  Our detailed
applications are to power-law galaxies with cores.  We have shown that
the caustics are then always simple closed curves, and have given
conditions for each of the four different caustic configurations.  We
have found an approximation for the magnitude of the central image
which applies throughout the region inside the large outer radial
caustic when the core is small and the slope index $2/B \leq 1$.  For
small cores and weaker cusps, that approximation is directly useful
only in a smaller inner region, but shows the magnitude of the central
image to grow on a short length scale.  We find that, for power-law
lenses with small cores and an inner tangential caustic, the sums over
separate pairs vary considerably with the image positions while the
signed sums over all four images are generally remarkably uniform.
That near-uniformity is a consequence of large cancellations between
terms which vary with the position of the source. Hence uncertainties
in the magnification can have major effects on estimates of those four
image sums.  This lessens their usefulness for the modeling of lenses
as proposed by Witt \& Mao (2000) and as used also by us in our Paper
I.  We have shown that similar large cancellations can arise with any
elliptically stratified potential, and not just power-laws.  Hence, they
will occur also with elliptically stratified densities in the limit of
low eccentricity, and perhaps more generally, though this needs to be
verified.

As an application, we have examined the constraints implied by the
missing central images of triplet and quintuplet systems. There is
only one convincing example of a gravitational lens system with a
central image, namely the triplet of the ultraluminous quasar
APM08279+5255. However, there are a total of $\sim 50$ doublets known.
Although this is a heterogeneous sample, discovered by different
observers using different techniques in different wavebands,
nonethless the probability of detecting central images does seem to be
very low. We take a central image as observable if the magnification
ratio of the faintest to the brightest image is $\gta 1 \%$ (although
some of the lens systems studied with the highest sensitivity by CLASS
do go deeper).  A rough summary of the observations is that the
probability of observing a central image for a triplet is $\lta 2
\%$. The absence of central images is understandable if the mass
distribution in the lensing galaxy population is nearly cusped, and
the cusp is isothermal or stronger. For typical source and lens
redshifts, the size of the core radius $a$ must be $\lta 300$ pc. The
slope of the gravitational potential $\beta$ must be $\lta 1$.

Most of the optical depth to strong lensing resides in the most
massive galaxies, namely giant ellipticals.  We know from high
resolution imaging of nearby giant ellipticals that they typically
have cuspy cores with $\beta \approx 0.7$ or $2/B \approx 1.3$ outside
the break radius of a few hundred parsecs (Faber et al. 1997). The
break radius corresponds to only a few tens of millarcseconds at a
typical lens redshift. Hence, strong lensing is primarily probing the
steep outer part of the cuspy core profile. This is much steeper than
isothermal, as the surface density is falling typically like
$r^{-1.3}$.  The cuspy cores by themselves can provide the explanation
of the missing central images.  Dark matter at large radii may alter
the slope of the projected mass distribution in the inner few
kpc. However, provided the slope remains isothermal or steeper and the
break radius remains small, then the central image remains
unobservable.  The ratio of the faintest to the brightest image in
cuspy core profiles is typically $\sim 0.001 \%$. Even the most
sensitive radio maps available probe only to a threshold of $0.05 \%$,
so this explains why the central images have so far remained missing
despite deeper searches. The sensitivity of the searches must be
increased by a factor of $\sim 50$ to find them.

\begin{acknowledgments}
NWE thanks the Royal Society for financial support.  The work of CH is
supported in part by NSF through grants DMS-9704615 and DMS-0104751.
We thank an industrious referee for a helpful report.

\end{acknowledgments}

\setcounter{section}{0}
\setcounter{equation}{0}

\begin{appendix}
\section{Caustics}
In this Appendix, we show that the power-law galaxies with cores in
the presence of external shear give rise to either one, three or five
images, depending on the position of the source and the extent of the
core radius. A point-like source is never lensed into more than five
images. The caustics are simple closed curves and there is exactly one
point on a caustic in each radial direction from the center of the lens.
The models differ from elliptically stratified density
distributions. For these, Witt \& Mao (2000) and Keeton, Mao \& Witt
(2000) showed that external shear can cause butterfly and swallowtail
cusps to develop on the caustics and so sextuple imaging and higher
does occur.

The caustics are curves in the source plane on which the imaging
equation (\ref{eq:newimaging}) has double roots.  The partial
derivative of the imaging equation with respect to $t$ vanishes at a
double root and hence we find caustics from the common solutions of
the two equations
\begin{eqnarray}
\label{eq:causticone}
t^B[L(t)-a^2K_1(t)]=-K_1(t), \\
\label{eq:caustictwo}
Bt^{B-1}[L(t)-a^2K_1(t)]+t^B[L^{\prime}(t)-a^2K_1^{\prime}(t)]
=-K_1^{\prime}(t).
\end{eqnarray}
Here, we have introduced the functions
\begin{equation}
K_1(t)=-(t-t_1)^2(t-t_2)^2, \qquad
L(t)= [(P_0-t)\zeta - Q \zetab][(P_0-t)\zetab -\Qb \zeta].
\end{equation}
Dividing the two sides of equations (\ref{eq:causticone}) and
(\ref{eq:caustictwo}) gives a sextic polynomial in $t$, whose
coefficients involve $\zeta$.  Alternatively, we can solve for $L(t)$
and $L^{\prime}(t)$ and divide the results to obtain a non-polynomial
equation, but one which is independent of $|\zeta|$ and depends only
on the angular argument $\phi$ of $\zeta$. This equation is
\begin{eqnarray}
           \label{eq:doubleroottwo}
           {B\over t}&=&(1-a^2t^B)\left[{K_1^{\prime}(t) \over K_1(t)}
                        -{L^{\prime}(t) \over L(t)}\right], \nonumber \\
&=&(1-a^2t^B)\left[
{-2P^3\!+\!6P^2|Q|\cos\theta\!-\!6P|Q|^2\!+\!2|Q|^3\cos\theta
                                      \over
(P^2\!-\!|Q|^2)(P^2\!+\!|Q|^2\!-\!2P|Q|\cos\theta)}\right].
\end{eqnarray}
As $a \rightarrow 0$, it reduces to equation (A2) of Paper I.  It
contains the complex angle $\theta$ defined by
$Q\barzeta/\zeta=|Q|e^{i\theta}$ rather than $\phi$. We find caustics
by searching for roots for $t$ along each angular direction in the
source plane. Once $t$ is known, $|\zeta|$ follows from equation
(\ref{eq:causticone}). In general, this search must be done
numerically, but one can deduce from the graph of the term in square
brackets (which has vertical asymptotes at $t=t_1$ and $t_2$ and is
plotted in Figure 7 of Paper I) and the extra $(1-a^2t^B)$ factor,
that there will always be a root in $(t_2,t_1)$ for a point on the
tangential caustic if $a<t^{-B/2}_2$, and another in $(t_1,a^{-2/B})$
for a point on the radial caustic if $a<t^{-B/2}_1$.  There are four
special directions for which points on the caustics can be found
explicitly. They are directions in which $Qe^{-2i\phi}$ is real, and
we now consider them in turn. 

When $Qe^{-2i\phi}=|Q|$ and $\cos\theta=1$, then $t=t_2$ at which $P=|Q|$
is a triple root of equation (\ref{eq:doubleroottwo}). It gives a cusp
at $|\zeta|=2|Q|t^{-B/2}_2\sqrt{1-a^2t^B_2}$ on the tangential caustic
provided $a<t^{-B/2}_2$. The other solution of equation
(\ref{eq:doubleroottwo}) is the root of $B(t-t_1)=2t(1-a^2t^B)$ in
$(t_1,a^{-2/B})$ for $a<t^{-B/2}_1$, which gives a point on the radial
caustic. The $t=t_2$ cusps on the
tangential caustic are naked if their value of
$|\zeta|=2|Q|t^{-B/2}_2\sqrt{1-a^2t^B_2}$ exceeds the value of
$|\zeta|$ for the fourth root for $t$ of equation
(\ref{eq:doubleroottwo}).  That value is given by equation
(\ref{eq:causticone}) evaluated for that root. The root cannot be
found explicitly when $a\not=0$.  Hence the marginal cases shown in
Fig.~\ref{fig:christhree} are found numerically by eliminating $t/t_1$
between the equations
\begin{eqnarray}
	\label{eq:nakedness}
	B(t-t_1)&=&2t(1-a^2t^B)\nonumber\\
	t^B(1-a^2t^B_2)(t_1-t_2)^2&=&t^B_2(1-a^2t^B)(t-t_1)^2,
\end{eqnarray}
and solving for $t_2/t_1$ for given $B$ and $a^2t^B_1$. The second of
equations (\ref{eq:nakedness}) comes from equating the two values
for $|\zeta|^2$. We can solve for $t$ explicitly in the coreless $a=0$
case, to give the condition 
\begin{equation}
        \label{eq:zeroanakedness}
	\frac{t_1}{t_2}>1+\frac{2}{B}
	\left[\frac{(B-2)t_2}{Bt_1}\right]^{(B-2)/2},
\end{equation}
for naked cusps.

The other special directions are those for which 
$Qe^{-2i\phi}=-|Q|$ and $\cos\theta=-1$. Then $t=t_1$
at which $P=-|Q|$ is a triple root of (\ref{eq:doubleroottwo}), and
gives a cusp point on a caustic at
$|\zeta|=2|Q|t^{-B/2}_1\sqrt{1-a^2t^B_1}$ provided $a<t^{-B/2}_1$.
The other solution of equation (\ref{eq:doubleroottwo}) satisfies
$B(t-t_2)=2t(1-a^2t^B)$. Such a solution exists provided $t_2
<a^{-2/B}$. It gives a point on the radial or tangential caustic
according to whether it is greater or less than $t_1$. 
The $t=t_1$ cusp is on the tangential caustic in the first
case, and on the radial caustic in the second, giving condition
(\ref{eq:twolips}) as that for which the double lips configuration
occurs. Double lips occur because the $t=t_1$ cusp always lies on
whichever is the inner caustic in this special direction; its
$|\zeta|$ value never exceeds that for the other root of equation
(\ref{eq:doubleroottwo}). The two $t$ and $|\zeta|$ values are equal 
only when $t=t_1$ is a quadruple root of equation (\ref{eq:doubleroottwo})
and then $B(t_1-t_2)=2t_1(1-a^2t_1^B)$. This latter case is the 
transitional one in which the two caustics share a cusp and
coincide locally, and marks the stage at which the region for
core triplets has shrunk to zero. 

A simple analytical approximation for the radial caustic can be found 
for $a$ small and $B \leq 2$. This caustic, which does not exist for $a=0$,
is large. It is found by looking for large roots of equations
(\ref{eq:causticone}) and (\ref{eq:caustictwo}) for $t$,
and approximating $K_1$ by $-t^4$ and $L$ by $t^2|\zeta|^2$.
Eliminating $t$ then gives the approximation
\begin{equation}
\label{eq:radcausticsmalla}
      |\zeta|^2=\xi^2+q^2\eta^2={B \over 2}\left[{ 2-B \over 2a^2}\right]
      ^{(2-B)/B}[1+O(a^{2/B})].
\end{equation}
Hence the radial caustic is approximately an ellipse elongated in the 
$\eta$-direction, as in Figs.~\ref{fig:christwo}b and d. 
The reason why those 
two ellipses for $a=0.1$ are not well described by equation
(\ref{eq:radcausticsmalla}) is that its limit of $\xi^2+q^2\eta^2=1$ is 
not accurate in the marginal case of $B=2$ until $a\ll 1$. A more
accurate formula for the radial caustic for $B=2$ is 
\begin{equation}
\xi^2+q^2\eta^2=1-3a^{2/3}\left[(1+\gamma_1)\xi^2+2q\gamma_2\xi\eta
                +q^4(1-\gamma_1)\eta^2 \right]^{2/3}+O(a^{4/3}).
\end{equation}

We showed in Paper I, by an analysis of the quartic obtained when
$a=0$ so there is no complicating $t^B$ power in equation
(\ref{eq:doubleroottwo}), that there is never more than one root in
$(t_2,t_1)$ and one in $(t_1,\infty)$. The additional $(1-a^2t^B)$
factor now present has only a small effect at finite $t$ when $a$ is
small. However, it does ensure that equation (\ref{eq:doubleroottwo})
always has a root in $(t_1,a^{-2/B})$ when this interval exists, which
is not the case when $a=0$ if $B\leq 2$. If more roots are to occur at
larger values of $a$, there must be transitional cases at which the
pair of equations (\ref{eq:causticone}) and (\ref{eq:caustictwo}) have
a multiple root. We show next, using the sextic derived from those
equations by eliminating the $t^B$ terms, that this cannot be, and
hence that the cases listed in the previous paragraph are the only ones
possible.

To study the roots for $t$ in $t>t_1$, we work with the variable
$\sigma=(t-t_1)/|Q|$, for which the sextic is:
\begin{eqnarray}
\label{eq:sextic}
{a^2 B|Q|^2 \over |\zeta|^2}\!\!\!\!\!\!
&&\sigma^3(\sigma^3+6\sigma^2+12\sigma+8)
+(B-2)\sigma^4+2[(B-2)(2+\cos\theta) \nonumber\\
&&-(p_0+\cos\theta)]\sigma^3+2(1+\cos\theta)[3(B-p_0-3)\sigma^2\\
&&+2(B-3p_0-5)\sigma-4(1+p_0)]=0,\nonumber
\end{eqnarray}
where $p_0=P_0/|Q|>1$. We let $C_n$ denote the coefficient of
$\sigma^n$ and apply Descartes' rule of signs (Henrici 1974) which
shows that there is just one positive root for $\sigma$ when there is
a single sign change in the coefficient sequence
$(C_6,C_5,C_4,C_3,C_2,C_1,C_0)$.  We take $(1 \pm \cos \theta)>0$
because the equations simplify in the special cases for which $\cos
\theta = \pm 1$ as discussed earlier and look for possible
sign changes in the coefficient sequence.  For $B>2$, $C_6$, $C_5$,
and $C_4$ are positive and $C_0$ is negative.  Also $C_2>C_1$ as in
Paper I, and $C_3>0$ if $C_2 \geq 0$.  For $B=2$, $C_6$, $C_5$, and
$C_4$ are positive, while $C_2$, $C_1$, and $C_0$ are negative.  For
$B<2$, $C_6$ and $C_5$ are positive, $C_2$, $C_1$, and $C_0$ are
negative, and $C_4>C_3$. In each case, only one sign change can and
does occur.

To show that there is just a single root in $t_2<t<t_1$, we re-express
the sextic in the variable $\tau=(t-t_2)/(t_1-t)$ which runs from 0 to
$\infty$. It is then
\begin{equation}
{8 a^2 B|Q|^2 \tau^3 \over |\zeta|^2} + (1+2\tau + \tau^2)
                   (c_4\tau^4+c_3\tau^3+c_1\tau+c_0)=0,
\end{equation}
where $c_4>0$, $c_3>0$, $c_1$, and $c_0<0$ are the coefficients of
equation (A4) of Paper I. Labelling this sextic as
$C_6\tau^6+C_5\tau^5+C_4\tau^4+C_3\tau^3+C_2\tau^2+C_1\tau+C_0=0$,
one finds that $C_6$, $C_5$, and $C_4$ are always positive
and $C_0$ always negative. For $B>2$, $C_2>C_1$ and $C_3>0$ if $C_2 \geq 0$.
For $B \leq 2$, $C_2$ and $C_1$ are both negative.
In either case, there is exactly one sign change in the coefficient
sequence, and therefore exactly one root for $t$ in $(t_2,t_1)$.

\section{Evaluation of the $R$ Coefficients}

In this Appendix, we show how to evaluate the coefficients
$R_1(t_1,t_2,a^2;j,\ell,k)$ and $R_2(t_1,t_2,a^2;j,\ell,k)$
of the powers $|\lambda|^{\ell}|\nu|^k$  introduced
in Section 3. They can always be expressed as finite sums.
We consider first the scale-free case for which we give compact explicit
expressions, and then the cored case for which we derive recursive
relations. We give explicit expressions for sums of coefficients $R_1+R_2$
needed for four-image sums in the scale-free case. Then we derive a
compact expression for the $R_1+R_2$ sums valid for any elliptically
stratified potential, and discuss its consequences.

\subsection{Scale-free Case}

Compact representations for the $R_1$ and $R_2$ terms can be derived
for the coreless $a=0$ case by partial differentiation using Leibniz's
rule. We find:
\begin{eqnarray}
	R_1(t_1,t_2,0;j,\ell,k)
	&=&-{1\over\ell!}{\partial^{\ell}\over\partial t^{\ell}_1}
	   \left[ {t_1^{Bj-1}\over(t_1-t_2)^{k+1}} \right]\nonumber\\
	&=&\sum\limits^{\ell}_{m=0}{jB-1\choose m}{t_1^{Bj-1-m}\over(\ell-m)!}
	   {(-1)^{\ell-m-1}(\ell+k-m)!\over k!(t_1-t_2)^{\ell+k+1-m}}.
\end{eqnarray}
This sum of $(\ell+1)$ terms can be written as the terminating
hypergeometric series
\begin{equation}
\label{eq:Ronehyper}
     R_1(t_1,t_2,0;j,\ell,k)=
	{(-1)^{\ell+1}t^{Bj-1}_1\over(t_1-t_2)^{\ell+k+1}}{\ell+k\choose\ell}
	\;_2F_1\left(-\ell,1-jB;-k-\ell;1-{t_2\over t_1}\right),
\end{equation}
with the proviso that, when $k=0$ and the hypergeometric series
becomes a geometric one, only the first $(\ell+1)$ terms of that
series are to be used. A similar expression for $R_2$ can be found
from it by using the transformation relation in
eq~(\ref{eq:Ronederiv}). It is
\begin{equation}
\label{eq:Rtwohyper}
     R_2(t_1,t_2,0;j,\ell,k)=
	{(-1)^{\ell}t^{Bj-1}_2\over(t_1-t_2)^{\ell+k+1}}{\ell+k\choose\ell}
	\;_2F_1\left(-k,1-jB;-k-\ell;1-{t_1\over t_2}\right).
\end{equation}

In obtaining separate formulae for the two pairs of bright images
rather than for all four, the present work extends that of Paper I for
the coreless $a=0$ case. We showed there that the coefficients
for the sums of four images can all be expressed in terms of
hypergeometric functions; that is we showed that
\begin{eqnarray}
\label{eq:hypersum}
     R_1(t_1,t_2,0;j,\ell,k)&+&R_2(t_1,t_2,0;j,\ell,k)
	={-1\over(\ell+k+1)!}\nonumber\\
	&\times&\prod\limits^{k+\ell+1}_{s=1}(jB-s) {t^{jB-k-1}_2\over
	t^{\ell+1}_1}\;_2F_1 \left(\ell+1,jB;\ell+k+2;1-{t_2\over
	t_1}\right).
\end{eqnarray}
We also showed in Appendix C of Paper I that the infinite
hypergeometric series (\ref{eq:hypersum})
can be represented as the sum of the two finite components
(\ref{eq:Ronehyper}) and (\ref{eq:Rtwohyper}).
We did not appreciate the significance of the two separate components
as two-image sums. We did warn of the tendency of
the separate $R_1$ and $R_2$ components to cancel for small
$(1-t_2/t_1)$, and that their
sum could be computed more easily using the rapidly convergent infinite
series (\ref{eq:hypersum}). A simple instance of this is the $\ell=0$
case for which
\begin{eqnarray*}
	R_1(t_1,t_2,0;j,0,k)&=&{-t^{Bj-1}_1\over(t_1-t_2)^{k+1}},\\
	R_2(t_1,t_2,0;j,0,k)&=&{t^{Bj-1}_2\over(t_1-t_2)^{k+1}}
	\sum\limits^k_{s=0}{jB-1\choose s}\left({t_1\over t_2}-1\right)^s.
\end{eqnarray*}
The sum here for $R_2$ consists of the first $(k+1)$ terms in the
infinite binomial expansion of $(t_1/t_2)^{Bj-1}$ in powers of
$(1-t_1/t_2)$; and hence the first $(k+1)$ terms in the expansion of
$-R_1$. Another instance occurs when $B=1$ or $B=2$. Then
$R_1+R_2\equiv 0$ for $j>0$ with the result that the sum of the four
signed magnifications are then independent of the position of the
source (Witt \& Mao 2000; see also Paper I). These examples are
simple instances of the large cancellations which can occur when
contributions of the direct and the inverted image pairs,
weighted with the signed magnifications, are combined.

\subsection{Cored Case}

The idea is to use Leibniz's rule to evaluate
\begin{equation}
	R_1(t_1,t_2,a^2;j,\ell,k)
	=-{1\over\ell!}{\partial^{\ell}\over\partial t^{\ell}_1}
    \left[{1\over(t_1-t_2)^{k+1}}{t^{Bj-1}_1\over(1-a^2t^B_1)^j}\right].
\end{equation}
The general derivative of the second factor is of the form
\begin{equation}
	{\partial^m\over\partial t^m_1}
	\left[{t^{Bj-1}_1\over(1-a^2t^B_1)^j}\right]
	={t^{Bj-m-1}_1\over(1-a^2t^B_1)^j}\sum\limits^m_{n=0}
	\alpha_{m,n}
	\left[{a^2t^B_1\over(1-a^2t^B_1)}\right]^n,
\end{equation}
where the coefficients $\alpha_{m,n}$, $m\geq n\geq 0$, can be found
iteratively from the relation
\begin{equation}
	\alpha_{m,n}=[B(j+n)-m]\alpha_{m-1,n}
	+B(j+n-1)\alpha_{m-1,n-1},\qquad
	\alpha_{0,0}=1.
\end{equation}
Use of Leibniz's rule then gives
\begin{eqnarray}
\label{eq:Roneasum}
     R_1(t_1,t_2,a^2;j,\ell,k)
	&=&{(-1)^{\ell+1}t^{Bj-1}_1\over
k!(t_1-t_2)^{\ell+k+1}(1-a^2t^B_1)^j}\\
	&\times&\sum\limits^{\ell}_{m=0}{(\ell+k-m)!\over m!(\ell-m)!}
	        \left({t_2\over t_1}-1\right)^m\sum\limits^m_{n=0}
			  \alpha_{m,n}\left[{a^2t^B_1\over
1-a^2t^B_1}\right]^n.\nonumber
\end{eqnarray}
In an analogous manner, we find;
\begin{eqnarray}
\label{eq:Rtwoasum}
     R_2(t_1,t_2,a^2;j,\ell,k)
	&=&{(-1)^{\ell}t^{Bj-1}_2\over
\ell!(t_1-t_2)^{\ell+k+1}(1-a^2t^B_2)^j}\\
	&\times&\sum\limits^{k}_{m=0}{(\ell+k-m)!\over m!(k-m)!}
	        \left({t_1\over t_2}-1\right)^m\sum\limits^m_{n=0}
			  \alpha_{m,n}\left[{a^2t^B_2\over
1-a^2t^B_2}\right]^n.\nonumber
\end{eqnarray}
The only $\alpha$'s for which it is simple to derive explicit
expressions are
\begin{equation}
	\alpha_{n,0}=\prod\limits^n_{s=1}(Bj-s),\qquad
	\alpha_{n,n}={B^n(j+n-1)!\over(j-1)!}.
\end{equation}
The first set are the only ones that appear in the $a=0$ case when
the double sums (\ref{eq:Roneasum}) and (\ref{eq:Rtwoasum}) reduce to
the hypergeometric sums (\ref{eq:Ronehyper}) and (\ref{eq:Rtwohyper})
respectively.

\subsection{Coefficient sums}

We begin by rewriting equation (\ref{eq:Rtwodefn}) for $R_2$ as
\begin{equation}
	\label{eq:Rtwogen}
	R_2( t_1,t_2,a^2;j,\ell,k )
	= \frac{1}{\ell!k!}\frac{\partial^{\ell}}{\partial t^{\ell}_1}
	  \frac{\partial^k}{\partial t^k_2}
	\left[-\frac{1}{2\pi i}\oint\nolimits_{\C_2}
	  \frac{G(t)\,dt}{(t-t_1)(t-t_2)} \right].
\end{equation}
The choice $G(t)=t^{Bj-1}/(1-a^2t^B)^j$ is needed for
equation (\ref{eq:Rtwodefn}), but a similar equation with a different
$G$ will arise for some other elliptically stratified potential
$\psi(\tau)$ with $t=2\psi^{\prime}(\tau)$.
We again choose $\C_2$ to be a contour which encloses
$t=t_2$ but not $t$ or any singularity of $G$. 
Then a residue calculation gives
\begin{equation}
	-\frac{1}{2\pi i}\oint\nolimits_{\C_2}
	   \frac{G(t)\,dt}{(t-t_1)(t-t_2)}
	=\frac{G(t_2)}{t_1-t_2}.
\end{equation}
There is a similar relation
\begin{equation}
	\label{eq:Ronegen}
	R_1(t_1,t_2,a^2;j,\ell,k)
	= \frac{1}{\ell!k!}\frac{\partial^{\ell}}{\partial t^{\ell}_1}
	  \frac{\partial^k}{\partial t^k_2}
	  \left[-\frac{1}{2\pi i}\oint\nolimits_{\C_1}
	  \frac{G(t)\,dt}{(t-t_1)(t-t_2)} \right],
\end{equation}
for the $R_1$ coefficient. When we evaluate its integral by residues,
and combine the results, we obtain
\begin{equation}
\label{eq:Rgensum}
R_1+R_2=-\frac{1}{\ell!k!}
         \frac{\partial^{\ell}}{\partial t^{\ell}_1}
         \frac{\partial^k}{\partial t^k_2} G[t_1,t_2]
       =-G[t_1,\ldots t_1,t_2,\ldots t_2].
\end{equation}
Here
\begin{equation}
	G[t_1,t_2]=\frac{G(t_1)-G(t_2)}{t_1-t_2},
\end{equation}
is a simple divided difference, while $G[t_1,\ldots t_1,t_2,\ldots t_2]$
has $t_1$ repeated $(\ell+1)$ times and $t_2$ repeated $(k+1)$ times,
and is a divided difference of order $(\ell+k+1)$ (Milne-Thomson
1960, chap. 1). 

The significance of formula (\ref{eq:Rgensum}) is as follows. 
Divided differences are
well-behaved when $(t_1-t_2)$ becomes small. More specifically, a mean
value theorem shows that
\begin{equation}
	R_1+R_2=-G[t_1,\ldots t_1,t_2,\ldots t_2]
	=-\frac{G^{(\ell+k+1)}(\bar{t})}{(\ell+k+1)!},
\end{equation}
where $\bar{t}$, at which the $(\ell+k+1)$th derivative of $G$ is
evaluated, is some value of $t$ in the interval $(t_2,t_1)$.  Thus,
when the axis ratio $q$ is close to 1 so that $(t_1-t_2)$ is small,
while the function $G$ varies smoothly in $(t_2,t_1)$ as in Section
B.1, then both $R_1$ and $R_2$ become very large when $(t_1-t_2)$ is
small because both contain negative powers up to and including
$(t_1-t_2)^{-\ell-k-1}$, while their sum varies little.  But this is
not the case with the lensing potential of Fig.~\ref{fig:christwo}a. 
Then $a$ is large and the negative $(1-a^2t^B)^j$ powers cause 
the derivatives of $G$ to be large, so that $R_1+R_2$ is then also large.

\end{appendix}

\end{document}